\documentclass[a4paper,fleqn,usenatbib]{mnras}
\usepackage{amsmath}
\usepackage{bm}
\usepackage{amssymb}
\usepackage{txfonts}
\usepackage{multicol}
\usepackage{subcaption}

\newcommand{\Ln}{\hat{\bm{L}}}
\renewcommand{\L}{\bm{L}}
\newcommand{\eps}{\varepsilon}
\newcommand{\abs}[1]{\left|#1\right|}

\usepackage[T1]{fontenc}
\usepackage{ae,aecompl}

\usepackage{graphicx}	
\usepackage{grffile}

\usepackage{epsf}
\usepackage{slashbox}
\usepackage{xspace}

\date{Accepted XXX. Received YYY; in original form ZZZ}

\pubyear{2025}
\AtBeginDocument{}%
\begin{document}
\label{firstpage}
\pagerange{\pageref{firstpage}--\pageref{lastpage}}
	\newcommand{\msun}{\,{\rm M}_{\odot}}
	\newcommand{\rsun}{R_{\odot}}
	\newcommand{\kms}{\, {\rm km\, s}^{-1}}
	\newcommand{\cm}{\, {\rm cm}}
	\newcommand{\gm}{\, {\rm g}}
	\newcommand{\erg}{\, {\rm erg}}
	\newcommand{\kel}{\, {\rm K}}
	\newcommand{\pc}{\, {\rm pc}}
	\newcommand{\kpc}{\, {\rm kpc}}
	\newcommand{\mpc}{\, {\rm Mpc}}
	\newcommand{\seg}{\, {\rm s}}
	\newcommand{\kev}{\, {\rm keV}}
	\newcommand{\hz}{\, {\rm Hz}}
	\newcommand{\etal}{et al.\ }
	\newcommand{\yr}{\, {\rm yr}}
	\newcommand{\gyr}{\, {\rm Gyr}}
	\newcommand{\eq}{eq.\ }
	\newcommand{\amunit}{\msun {\rm AU^2/yr}}
	\def\arcsec{''\hskip-3pt .}
    \def\half{{\textstyle{\frac{1}{2}}}}
    \def\gothg{\,\mathfrak G}
    \def\ffrac#1#2{{\textstyle\frac{#1}{#2}}}
    \newcommand{\kmsf}{\frac{\rm km}{\rm s}}
    \newcommand{\Myr}{\,{\rm Myr}}
    \newcommand{\Gyr}{\,{\rm Gyr}}
    \newcommand{\D}{\mathrm{d}}
    \newcommand{\Msun}{\,\mathrm{M}_{\odot}}
    \newcommand{\E}{\mathrm{VRR}}
    \newcommand{\tot}{\mathrm{tot}}
    \newcommand{\J}{\mathcal{J}}
    \renewcommand{\L}{\bm{L}}
    \renewcommand{\O}{\mathbf{O}}
    \newcommand{\G}{\gothg}
    \newcommand{\In}{\rm in}
    \newcommand{\Out}{\rm out}
    \newcommand{\rms}{\rm RMS}
    \renewcommand{\r}{\bm{r}}
    \newcommand{\sat}{\mathrm{sat}}
    \newcommand{\vrr}{\mathrm{vrr}}	
    \newcommand{\IMBH}{\bullet}
    \newcommand{\MSMBH}{M}    
    \newcommand{\orb}{\rm orb} 
	\def\gapprox{\;\rlap{\lower 3.0pt                       
			\hbox{$\sim$}}\raise 2.5pt\hbox{$>$}\;}
	\def\lapprox{\;\rlap{\lower 3.1pt                       
			\hbox{$\sim$}}\raise 2.7pt\hbox{$<$}\;}
	
	\newcommand{\figsizeFour}{9.0cm}
	
	\newcommand{\figwidthSingle}{14.0cm}
	\newcommand{\figwidthDouble}{7.50cm}
	
	\newcommand{\figbig}{\figwidthSingle}
	\newcommand{\figsmall}{\figwidthDouble}

	\newcommand{\figappend}{\figwidthSingle}
	
	\newcommand{\reqOne}[1]{equation~(\ref{#1})}
	\newcommand{\reqTwo}[2]{equations~(\ref{#1}) and~(\ref{#2})}
	\newcommand{\reqNP}[1]{equation~\ref{#1}}
	\newcommand{\reqTwoNP}[2]{equations~\ref{#1} and~\ref{#2}}
	\newcommand{\reqTo}[2]{equation~(\ref{#1})-(\ref{#2})}
	\newcommand{\rn}[1]{(\ref{#1})}
	\newcommand{\ern}[1]{equation~(\ref{#1})}
	\newcommand{\be}{\begin{equation}}
	\newcommand{\ee}{\end{equation}}
	\newcommand{\ff}[2]{{\textstyle \frac{#1}{#2}}}
	\newcommand{\ben}{\begin{enumerate}}
		\newcommand{\een}{\end{enumerate}}

\newcommand{\nring}{\texttt{N}-Ring\xspace}
\newcommand{\phicpu}{$\phi$-CPU\xspace}
\newcommand{\bk}[1]{{\color{red}#1 }}
\newcommand{\tp}[1]{\textcolor{blue}{$\langle$\textbf{TP: #1}$\rangle$}}
\newcommand{\byg}[1]{\textcolor{cyan}{$\langle$\textbf{BYG: #1}$\rangle$}}
\newcommand{\bg}[1]{\textcolor{cyan}{#1}}

\title[Angular-momentum pairs in spherical systems]{Angular-momentum pairs in spherical systems: applications to the Galactic centre}

\author[Panamarev, Ginat \& Kocsis]
  {Taras Panamarev$^{1,2}$\thanks{Corresponding author email: panamarevt@gmail.com}, Yonadav Barry Ginat$^{1,3}$ and Bence Kocsis$^{1,4}$ \\
   \\
      $^1$ Rudolf Peierls Centre for Theoretical Physics, Parks Road, OX1 3PU, Oxford, United Kingdom\\
      $^2$ Fesenkov Astrophysical Institute, Observatory 23, 050020 Almaty, Kazakhstan \\
      $^3$ New College, Holywell Street, Oxford, OX1 3BN, United Kingdom \\ 
      $^4$ St.~Hugh's College, St.~Margaret's Road, Oxford, OX2 6LE, United Kingdom
      }

\maketitle

\begin{abstract}
Consider a system of point masses in a spherical potential. In such systems objects execute planar orbits covering two-dimensional rings or annuli, represented by the angular-momentum vectors, which slowly reorient due to the persistent weak gravitational interaction between different rings. This process, called vector resonant relaxation, is much faster than other processes which change the size/shape of the rings. The interaction is strongest between objects with closely aligned angular-momentum vectors. In this paper, we show that nearly parallel angular-momentum vectors may form stable bound pairs in angular-momentum space. We examine the stability of such pairs against an external massive perturber, and determine the critical separation analogous to the Hill radius or tidal radius in the three-body problem, where the angular-momentum pairs are marginally disrupted, as a function of the perturber's mass, the orbital inclination, and the radial distance. 
Angular-momentum pairs or multiples closer than the critical inclination will remain bound and evolve together in angular-momentum-direction space under any external influence, such as anisotropic density fluctuations, or massive perturbers. This study has applications in various astrophysical contexts, including galactic nuclei, in particular the Milky Way's Galactic centre, globular clusters, or planetary systems. In nuclear star clusters with a central super-massive black hole, we apply this criterion to the disc of young, massive stars, and show that clusters in angular-momentum space may be used to constrain the presence of intermediate-mass black holes or the mass of the nearby gaseous torus. 
\end{abstract}


\section{Introduction}\label{sec:INTRO}

The general $N$-body problem of self-gravitating systems does not have a closed-form solution; however, when the system has certain symmetries or a hierarchical structure with well-separated time-scales, the long-time dynamics become tractable and amenable to detailed study \citep[see, e.g.][for a review of the $N$-body problem]{HeggieHut2003}. In these systems, the underlying assumption is that the short-term orbital motions can be averaged out -- this is the basis of the orbit-averaging method in Hamiltonian perturbation theory \citep[e.g.][]{Arnoldetal2006}. By averaging over the rapid orbital period, one obtains a simplified ``secular'' Hamiltonian that governs the long-term evolution of the system’s orbital elements, such as the orbital energy and angular momentum.

Several astrophysical systems allow for such a simplification, including planetary systems, galactic discs, and stellar clusters, such as globular clusters and nuclear star clusters. In the case of star clusters, they are approximately spherically symmetric, implying that to first order (i.e.~the monopole in a multipole expansion) the stars in them move on planar orbits with conserved orbital energy and angular momentum in the smooth mean-field gravitational potential of the system. Over longer time-scales, the orbital energy and angular-momentum vectors slowly diffuse due to the fluctuating component of the potential or close two-body encounters \citep{Binney}. A system of $N$ bodies bound to a massive central body, including planets around a star in a planetary system or stars around a super-massive black hole (SMBH) in a nuclear star cluster, evolves similarly since the Keplerian potential is spherically symmetric. In nuclear star clusters, all objects perform Keplerian eccentric orbits around the central SMBH with nearly fixed period ($\sim 10$--$10^4\,\mathrm{yr}$ in the case of the Milky Way, e.g. \citealt{Kocsis2011}), semi-major axis $a$, eccentricity $e$, orbital inclination $i$, and argument of ascending node $\Omega$, corresponding to fixed orbital energy $E$ and angular-momentum vector $\L$. These eccentric orbits additionally execute rapid in-plane apsidal precession, due to the mean-field potential of the stellar distribution on time-scales of $\sim 10^4\,\mathrm{yr}$, longer than the orbital period, covering two-dimensional rings (or annuli) between their respective periapsis and apoapsis. On even longer time-scales, the direction of the orbital planes or angular-momentum vectors reorient ($\sim 10^6\,\mathrm{yr}$) \citep{Rauch1996,Kocsis2015}. The eccentricities and semi-major axes change on average even more slowly ($\sim 10^8\,\mathrm{yr}$ and $\sim 10^9\,\mathrm{yr}$, respectively).

This time-scale hierarchy allows one to employ a double orbit-averaging procedure: first, one averages over the rapid orbital period, replacing the time-varying orbit with a stationary eccentric wire, and then one also averages over the longer apsidal-precession period induced by the smooth spherical stellar potential, replacing the planar-precessing orbit with rings or two-dimensional annuli, whose shapes are preserved (because semi-major axes and eccentricities are conserved over the relevant time-scales). The gravitational torques between these rings drive a slow change of their orientations; the relevant dynamical variables are thus the directions of orbital angular momenta, $\hat{\L}$, which are set by the inclination and the argument of the ascending node. This formalism is referred to as ``vector resonant relaxation'' (VRR) \citep{Rauch1996,Kocsis2011,Kocsis2015,Bar-Or2018,Fouvry+2022,Panamarev2022}. The VRR equations of motion can be solved exactly for the case of two rings; this results in a precession of the angular-momentum vectors of the two bodies about their total angular momentum, with a frequency which depends on the relative orbital configuration of the bodies \citep{Kocsis2015}. 

In this paper, we show that nearly parallel angular momentum vectors may form stable bound pairs in angular-momentum space. We examine the stability of such pairs against an external massive perturber, and determine the critical separation (in angular-momentum space) where the angular-momentum pairs are marginally disrupted, as a function of the perturber's mass, orbital inclination, and radial distance from the centre of the spherical potential. This is completely analogous to the Hill radius, or tidal radius, in the restricted three-body problem, where a binary may be split by a massive perturber at the distance where the tidal force of the perturber equals the force exerted by the binary companion \citep{Hill1887,Hills1988,Arnoldetal2006}. Note, that the angular-momentum pairs studied in this paper do not form physical binaries in terms of separation or orbital energy, as they are both orbiting around the centre of the cluster and they may be very distant in physical space; nevertheless, they may be close in angular-momentum direction space if their orbits have a low mutual inclination -- they are then bound to each other, due to their large negative mutual VRR energy. These angular-momentum pairs (or angular-momentum clusters if more than two objects are bound together by VRR) may then execute a correlated random walk in angular-momentum-direction space due to massive perturbers or the fluctuating anisotropy of the cluster.

This problem is directly relevant to galactic nuclei: in the Milky Way, which hosts a super-massive black hole of mass $\MSMBH = 4.3\times10^6\msun$ \citep{Ghez2000, Genzel2010, Gravity2024}, the nuclear star cluster (NSC) is dominated by old stars but also contains a population of young stars, some of which reside in a disc-like configuration \citep[e.g.][]{LevinBeloborodov2003,PaumardEtAl2006,vonFellenberg2022,Siyao2023}. Initial evidence for this disc arose from the clustering of stellar angular-momentum vector directions $\hat{\L}$ \citep{LevinBeloborodov2003}; however, recent analyses reveal a more complex distribution featuring multiple sub-clusters of these vectors \citep{Bartko2009,vonFellenberg2022}, on the sphere of $\hat{\L}$. Although certain sub-structures remain under debate \citep{Siyao2023}, there is broad agreement on at least two distinct angular momentum clusters,
one of which is retrograde relative to the other \citep{vonFellenberg2022,Siyao2023}. A natural formation scenario invokes the fragmentation of a past gaseous accretion disc \citep{LevinBeloborodov2003}, but the presence of retrograde stars suggests additional processes -- such as perturbations by an external massive perturber or torques from a rotating NSC \citep[cf.][]{Levin2024}. The young stellar disc is estimated to be $\sim 6$ Myr old \citep{Genzel2010,Habibi2017}, comparable to the VRR time-scales at these radii \citep{Kocsis2011}. Consequently, resonant relaxation has been widely invoked to explain the observed distribution of stellar angular-momentum vectors under the influence of the NSC \citep{Kocsis2011,Kocsis2015,Roupas2017,Giral2020,Fouvry2019,Fouvry+2022,Magnan2021,Panamarev2022,Wang_Kocsis2023,FloresFouvry2024}, as well as their modification by an external perturber \citep{Szolgyen2021,Ginat2022,Fouvryetal2023,Levin2024}. Although VRR is a promising mechanism for producing the observed misalignments, other relaxation processes -- such as scalar resonant relaxation \citep{Bar-Or2018} and two-body (non-resonant) relaxation -- may also contribute to the dynamics (see e.g.~\citealt{Panamarev2022}, who showed that two-body relaxation dominates for galactic nuclei with a light SMBH). The most accurate way to treat the simultaneous action of all of these processes is with high-resolution direct $N$-body simulations, which integrate the exact gravitational forces without relying on the orbit-averaged VRR approximation; however, direct $N$-body simulations of the full NSC remain computationally prohibitive -- state-of-the-art models of the entire nuclear cluster require months to complete, even on modern GPUs \citep{Panamarev2019, Panamarev2022}. 

We therefore aim, in this paper, to develop a simple semi-analytical framework to determine when initially-coupled angular-momentum vectors become decoupled by a massive perturber. This model can be used to (i) obtain order-of-magnitude constraints on the perturber’s mass and orbital parameters at the Galactic centre, (ii) guide the set-up of computationally demanding direct $N$-body simulations, and (iii) allow one to study the theoretical stability of angular-momentum binaries. 

In this paper, we study the conditions under which two mutually torquing rings are driven to decoupling by an external perturber. The paper is organised as follows: in Sec.~\ref{sec:VRRformulation} we revisit the VRR formalism, describe the set-up of the problem and derive some analytical expressions. We then use them to derive a stability criterion in Sec.~\ref{sec: rel_torques} -- our main result in this paper -- which we test extensively with simulations (described there). We comment on the validity of the VRR-derived criteria in Sec.~\ref{sec:validity}, discuss astrophysical applications of this problem in Sec.~\ref{sec:applications_to_GC}, and summarise the paper in Sec.~\ref{sec:summary}.
We provide details of analytical calculations and numerical simulations in the appendices.

\section{Secular VRR Hamiltonian and Asymptotic Limits}
\label{sec:VRRformulation}

We begin by considering a system of \(N\) objects moving on apsidally precessing Keplerian orbits around a central SMBH, with fixed semi-major axes \(a_i\) and eccentricities \(e_i\). The central mass is denoted \(\MSMBH\). We assume that \(\sum_i m_i \ll \MSMBH\), so the gravitational forces from the stars on each other can be treated as small perturbations on otherwise Keplerian orbits. General relativistic corrections and external torques (e.g.\ from a massive perturber or frame-dragging) are neglected or assumed sub-dominant on the time-scales of interest, unless stated otherwise. Within the VRR framework, over time-scales long compared to an orbital period but short compared to changes in the scalar angular momenta \(L_i\), the orbits can, therefore, be treated as apsidally precessing annuli whose mutual gravitational interaction is suitably averaged \citep[see e.g.][]{Kocsis2015}. To build an intuitive understanding of these interactions, we begin by analysing the mutual torques between two circular rings; we then describe the Hamiltonian of the system, which provides a compact and general framework for describing their long-term secular evolution.

\subsection{Torques between rings}
\label{subsec:2rings}

In near-Keplerian potentials over extended time-scales, stellar orbits are effectively approximated as `stellar rings'. These rings change in shape and orientation due to perturbations to the Keplerian potential -- their mutual gravitational interactions. To study the effect of perturber initially precessing pair of stars, let us start with a simple concept of two rings exerting mutual torques. 

Following \citet{Nayakshin2005}, we analyse the gravitational interaction between two circular rings in a galactic nucleus. Consider ring 1 with radius $R_1$ and ring 2 with radius $R_2$. These rings are inclined at an angle $I$ to each other. We approach this problem using two coordinate systems, denoted as $x,y,z$ for ring 1 and $x',y',z'$ for ring 2, both centred on the SMBH. Ring 1 is aligned with the plane defined by $z=0$, while ring 2 corresponds to the plane where $z'=0$. The inclination angle $I$ is thus the angle formed between the $z$ and $z'$ axes, and the axes $x$ and $x'$ are aligned along their mutual line of nodes. 

The gravitational torque exerted by ring 2 on ring 1 can be expressed as \citep{Nayakshin2005}:
\begin{equation}
\bm{\tau}_{12} = G \eta_1 \eta_2 R_1 R_2 \int_0^{2\pi} \mathrm{d}\phi_1 \int_0^{2\pi} \mathrm{d}\phi_2 \frac{[\bm{r}_1 \times \bm{r}_2]}{|\bm{r}_2 - \bm{r}_1|^{3}}
\label{eq:tau1}
\end{equation}
where $\eta_1 = m_1/2\pi R_1$ and $\eta_2 = m_2/2\pi R_2$ are the linear mass densities of ring 1 and ring 2, respectively. Here, $m_1$ and $m_2$ are the total masses of rings 1 and 2, respectively. The integrals are taken over $\phi_1$ and $\phi_2$, which represent the azimuthal angles in the coordinate systems corresponding to ring 1 and ring 2. This equation reveals that co-planar rings (\(I = 0\)) exert no gravitational torque on each other because the configuration is symmetric. Similarly, for polar orbits, when \(I = \pi/2\), the integral cancels out due to equal and opposite contributions from points on opposite sides of ring 2; the system's symmetry also leads to $\left(\tau_{12}\right)_z = 0$.
Thus, if we now take the rings to represent a star and an IMBH, with \(m_2 \gg m_1\), the \(x\)-component of the torque induces nodal precession of the star's angular momentum vector around that of the IMBH.

While Eq.~\eqref{eq:tau1} offers an intuitive understanding of the gravitational interactions between two rings, extending this analysis to the $N$-ring system as a whole, and generalising for elliptical orbits, are more conveniently achieved through the Hamiltonian formalism. The gravitational interaction-energy of the ring $i$ with all other rings \citep{Kocsis2015} can be written as:
\begin{equation}\label{eq:H_VRR}
 H_{i} = -\sum_{j=1, i < j}^{N}\sum_{\ell=0}^{\infty} \J_{ij\ell}\,P_{\ell}\big( \Ln_{i}\cdot \Ln_{j} \big),
\end{equation}
where $P_{\ell}$ denotes the Legendre polynomial of degree $\ell$, the unit vectors $\Ln_{i} \equiv \L_{i}/L_{i}$ are perpendicular to the orbital planes, and, as above, $N$ is the total number of rings in the system. The term $\J_{ij\ell}$ represents the coupling coefficients, defined as
\begin{equation}\label{eq:Jijl}
\J_{ij\ell} = \frac{ G m_i m_j}{a_{\Out}}
 [P_{\ell}(0)]^2\, s_{ i  j \ell}\, \alpha_{ij}^{\ell}\,,
\end{equation}
where $\alpha_{ij} \equiv a_{\mathrm{in}} / a_{\mathrm{out}}$, $a$ is the semi-major axis with indices `${\mathrm{out}}$' and `${\mathrm{in}}$' denoting the larger and smaller semi-major axes among $i$ and $j$, respectively; the coefficient $s_{i j \ell}$ is:
\begin{align}\label{eq:s_ijl}
 s_{ij\ell} =
\int_0^{\pi} \frac{\D \phi}{\pi^2} \int_0^{\pi} \D \phi'
\frac{\min\left[(1 + e_{\In}\cos\phi)\,,\; \alpha_{ij}^{-1}(1 + e_{\Out} \cos\phi')\right]^{\ell+1}
}{ \max\left[ \alpha_{ij}(1 + e_{\In} \cos\phi)\,,\; (1 + e_{\Out}  \cos\phi')\right]^{\ell}}\,.
\end{align}
The couplings $\mathcal{J}_{ij\ell}$ are constant, for they are independent of the dynamical variables $\hat{\L}_i$ -- namely, the inclination and the argument of the ascending node.
For non-overlapping eccentric orbits $s_{ij\ell}$ simplifies to:
\begin{equation}
s_{ij\ell} = \frac{\chi_{ \Out }^{\ell}}{\chi_{ \In }^{\ell+1}}
 P_{\ell+1}(\chi_{ \In })P_{\ell-1}(\chi_{ \Out })\quad {\rm
   if}~r_{a,\In}<r_{p,\Out}\,,
\label{e:xxxyyy}
\end{equation}
where $b_i=a_i\sqrt{1-e_i^2}$, $\chi_i=a_i/b_i$, and $r_{\rm a}$ and $r_{\rm p}$ denote the apocentre and pericentre, respectively. The subscript `in' refers to the ring with the smaller semi-major axis (and similarly for `out'). 
For circular orbits, as $e_{\In} = e_{\Out} = 0$, the coefficient satisfies $s_{ij\ell}=1$. The Legendre polynomials at zero (in terms of the gamma function) are given by
\begin{equation} \label{eq:pnzero}
P_{2n}(0) = (-1)^n\frac{\Gamma(2n+1)}{2^{2n}[\Gamma(n+1)]^2}, \quad P_{2n+1}(0) = 0.
\end{equation}

Note that the Hamiltonian \eqref{eq:H_VRR} is double orbit-averaged: over the orbital period \( t_{\orb} \) and the apsidal precession period \( t_{\mathrm{prec}} \gg t_{\orb} \). As a result, semi-major axes and eccentricities of the orbits are conserved, leading to the conservation of the angular-momentum magnitudes, but the directions of the angular-momentum vectors may change. This approach is justified for the time-scales of interest: \( t_{\mathrm{prec}} \ll t \ll t_{\mathrm{rel}} \), where \( t_{\mathrm{rel}} \) is the two-body relaxation time.

\subsection{Equation of motion and asymptotic limits}
\label{subsec:asymp}

From the Hamiltonian \eqref{eq:H_VRR}, one can get the equation of motion \citep{Kocsis2015}:
\begin{equation}
\label{eq:EOM}
    \frac{\mathrm{d}\L_i}{\mathrm{d}t} = \mathbf{\Omega}_i \times \L_i = - \sum_{j\ell} \frac{\J_{ij\ell}}{L_i L_j} P'_{\ell}(\Ln_i\cdot \Ln_j) \L_j \times \L_i\,.
\end{equation}
The equation represents a rotation of the angular-momentum vector $\L_i$ with frequency $\boldsymbol{\Omega}_i$ (which generically depends on all the $\L_j$'s and on time). Here quadrupole, \(\ell=2\), terms dominate the torques when the orbits are well separated ($\alpha_{ij} \ll 1$); in particular, for circular orbits in the quadrupole approximation, the angular frequency, due to the pairwise torque, is:
\begin{align}
\label{eq:quad_circ}
\Omega_{i,j} &= 2\pi \omega_i \frac{m_{j}}{\MSMBH} \left(\frac{r_i}{r_\mathrm{out}}\right)\left(\frac{r_\mathrm{in}}{r_\mathrm{out}}\right)^2 [P_2(0)]^2P'_2(\cos{\theta}) 
\nonumber\\
&=
\frac{3\pi}{2}\omega_{i}\,
\frac{m_{j}}{\MSMBH}\,
\cos\theta\;
\begin{cases} 
\left( \dfrac{r_i}{r_j} \right)^3 & \text{if } r_i < r_j \\[1.2em] 
\left( \dfrac{r_j}{r_i} \right)^2  & \text{if } r_i \geq r_j 
\end{cases},
\end{align}
$\MSMBH$ is the mass of the SMBH, $\Omega_{i,j}\neq\Omega_{j,i}$, but $\Omega_{i,j}\times L_i = \Omega_{j,i}\times L_j$, and $\omega_i=(G\MSMBH/a_i^3)^{1/2}$ is the orbital frequency around the SMBH and $\theta$ is the angle between the angular-momentum vectors given by $\cos\theta = \hat{\L}_i \cdot \hat{\L}_j$. This expression is a good approximation when $r_{\In}/r_{\Out} \lesssim 0.3$, as we show in appendix \ref{app:energy}. 
One can use an asymptotic expansion (as $\ell \to \infty$) to derive an accurate approximation for arbitrary $r_{\In}/r_{\Out}$; we do so explicitly in Appendix~\ref{app:energy}. 

Moving to eccentric orbits, for non-overlapping ones, where the periapsis of the outer orbit is outside the
apoapsis of the inner orbit, we can also use the quadrupolar approximation for radially large orbital separations, finding
\begin{align}
\Omega_{i,j} &= \frac{3\pi}{2}\omega_{i}\frac{m_{j}}{\MSMBH}\frac{a_{i}a_{\rm in}^{2}}{a_{\rm out}^3 }\frac{\chi_{j}\chi_{\rm out}^3P_{3}(\chi_{\rm in}) }{\chi_{\rm in}^{3}}\cos\theta 
\nonumber\\
&= 
\frac{3\pi}{2}\omega_{i}\,
\frac{m_{j}}{\MSMBH}\,
\cos\theta\;
\begin{cases}
\displaystyle
\left(\frac{a_i}{a_j}\right)^{3}
\left(\frac{\chi_j}{\chi_i}\right)^{3}\;
\chi_j\,P_{3}(\chi_i),
& a_i < a_j,\\[1.2em]
\displaystyle
\left(\frac{a_j}{a_i}\right)^{2}
\left(\frac{\chi_i}{\chi_j}\right)^{2}\;
\chi_i\,P_{3}(\chi_j),
& a_i \ge a_j.
\end{cases}
\label{eq:quad_nonover}
\end{align}
On the other hand, for orbits with close separations which include overlapping or marginally overlapping orbits, large $\ell$ harmonics dominate, because the coefficient $\mathcal{J}_{ij\ell}$ decays slowly with $\ell$, and the torques can be approximated by asymptotic expressions for $\mathcal{J}_{ij\ell}$ at $\ell \to \infty$. As we show in Appendices~\ref{App:Legendre} and~\ref{app:asymp} \citep[cf.][]{Kocsis2015} an asymptotic expression valid for both overlapping ($z=1$) and non-overlapping orbits ($z<1$) is 
\begin{align}
    \mathbf{\Omega}_{i,j}  =& -  k_{ij} \omega_{\rm orb,i} 
     \left\{
     \frac{1/2}{1-x}\ln\left[\frac{(1+X+z)(1+Y-z)}{4}\right]
        \right.\nonumber\\
        &\quad-\left.\frac{1/2}{1+x}\ln\left[\frac{(1+X-z)(1+Y+z)}{4}\right]
     \right\}\Ln_j\,,         
\end{align}  
where $x \equiv \cos \theta$, $X\equiv (1-2 xz + z^2)^{1/2}$, $Y\equiv (1+2 xz + z^2)^{1/2}$ and
\begin{equation}
    z = 
        \left\{
        \begin{array}{ll}
        \dfrac{r_{\rm a,in}}{r_{\rm p,out}} & {\rm if~nonoverlapping}\,,  \\[2ex]
        1 & {\rm if~overlapping}\,.
    \end{array}
    \right.
\nonumber
\end{equation}
For overlapping orbits this simplifies as 
\begin{align}
    \mathbf{\Omega}_{i,j}  &= -  k_{ij} \omega_{\rm orb,i} 
     \left\{\frac{\ln[(1+s)c]}{4s^2}
        -\frac{\ln[(1+c)s]}{4c^2}
        \right\}\Ln_j \nonumber\\&\approx - \frac{1}{2} k_{ij} \omega_{\rm orb,i}\cot \theta\,\Ln_j \,.
\end{align}  
where $s \equiv \sin (\theta/2)$, $c \equiv \cos (\theta/2)$ and
\begin{align}     
    k_{ij}&= \frac{\J_{ij}}{L_i\omega_i} \nonumber\\&=    
    \left\{
        \begin{array}{ll}
        \dfrac{m_{j}a_i}{\pi^2 \MSMBH r_{\rm p,out}} 
    \dfrac{[(1+e_{\rm in})(1-e_{\rm out})]^{3/2}}{(e_{i}e_{j})^{1/2}(1-e_i^2)^{1/2}}  &\hspace{-2pt}\text{if non-overlapping}\,\\[2ex]
        \dfrac{4}{\pi^3} \dfrac{m_{j}}{\MSMBH} \dfrac{I_2}{a_j(1-e_i^2)^{1/2}} & \hspace{-2pt}\text{if overlapping}
    \end{array}
    \right.
\end{align}
Here,  $I_2\equiv I^{(2)}(r_{\rm p,in},r_{\rm p,out},r_{\rm a,in},r_{\rm a,out})$ is an analytic function expressed with elliptic integrals in a closed form, given by Eq.~(B68) of \cite{Kocsis2015} which simplifies for the case of a circular ring embedded within an elliptical one, as we show in Appendix \ref{app:I2}. In Appendix~\ref{app:EOM-corr} we show how to construct equations of motion valid for arbitrary values of radial separation of the orbits. 

In \S\,\ref{sec: rel_torques} below, we will use these asymptotic expansions, together with exact low-\(\ell\) terms, to derive and compare analytical torque-estimates in systems influenced by an additional massive perturber. 

\section{Relative torques}
\label{sec: rel_torques}

\begin{figure*}
  \centering
  \begin{subfigure}[b]{0.48\linewidth}
    \centering
    \includegraphics[width=\linewidth]{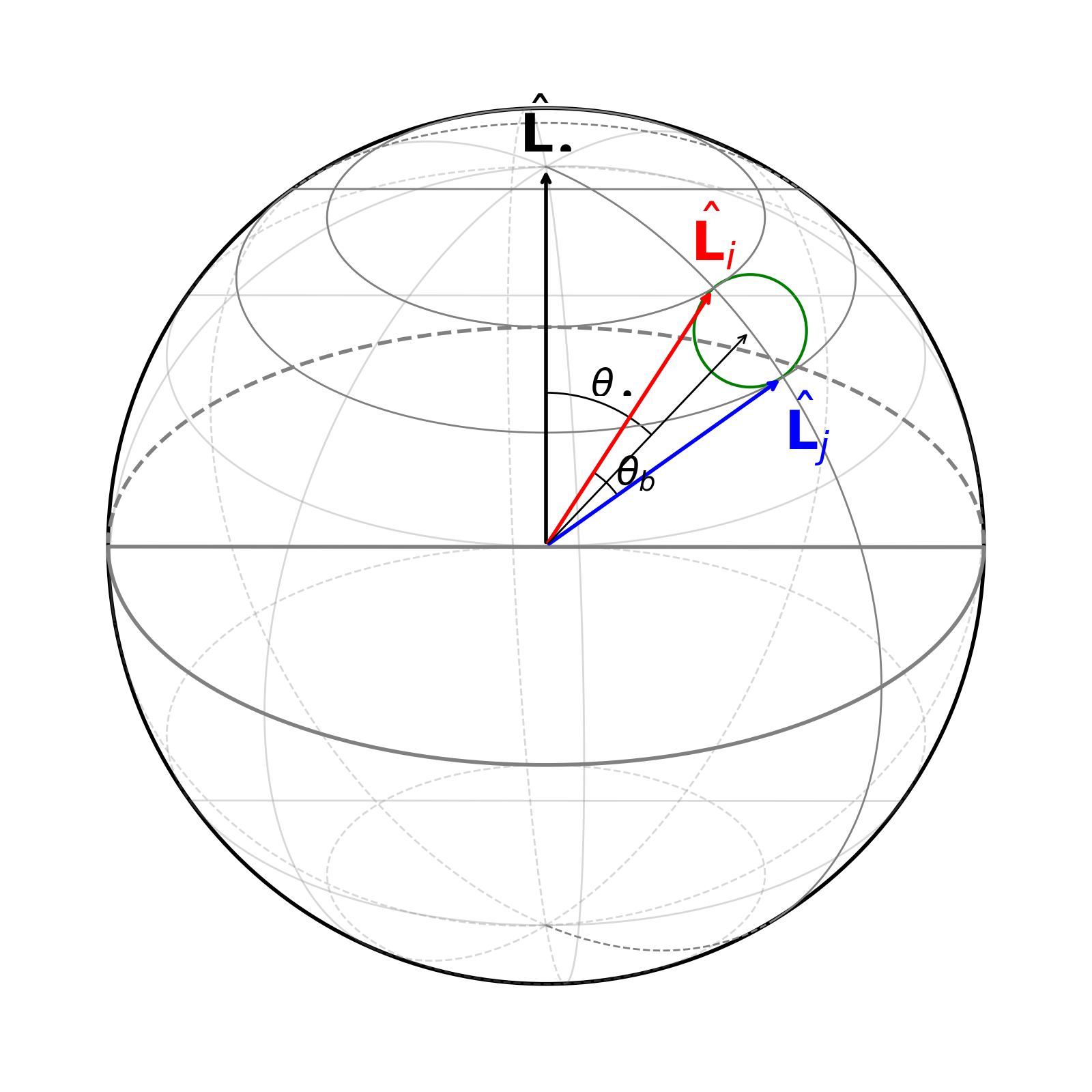}
    \caption{Initial set-up} 
    \label{fig:set-up}
  \end{subfigure}\hfill
  \begin{subfigure}[b]{0.46\linewidth}
    \centering
    \includegraphics[width=\linewidth]{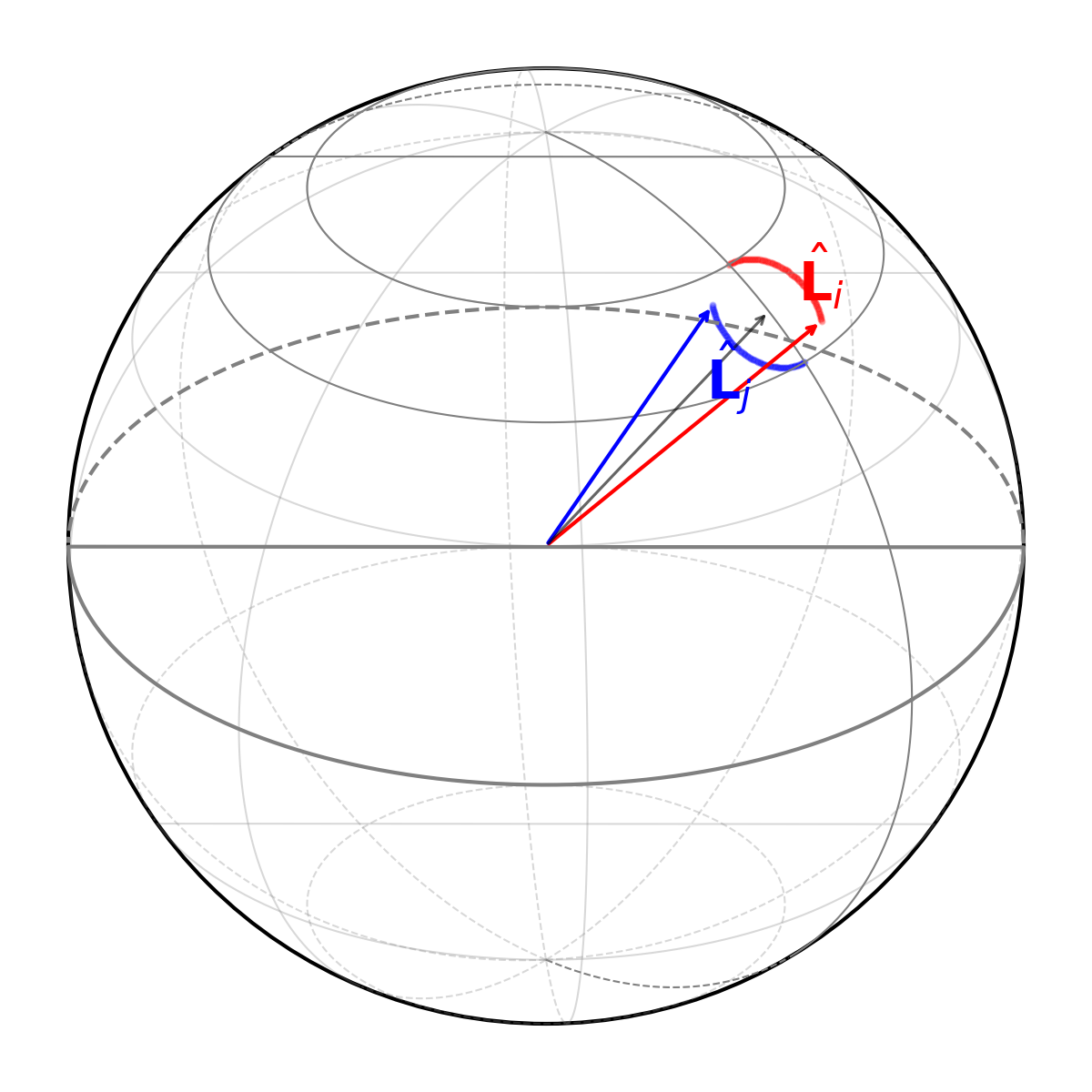}
    \caption{Precession in the absence of the IMBH}
    \label{fig:noIMBH}
  \end{subfigure}

  \vspace{0.8ex}

  \begin{subfigure}[b]{0.46\linewidth}
    \centering
    \includegraphics[width=\linewidth]{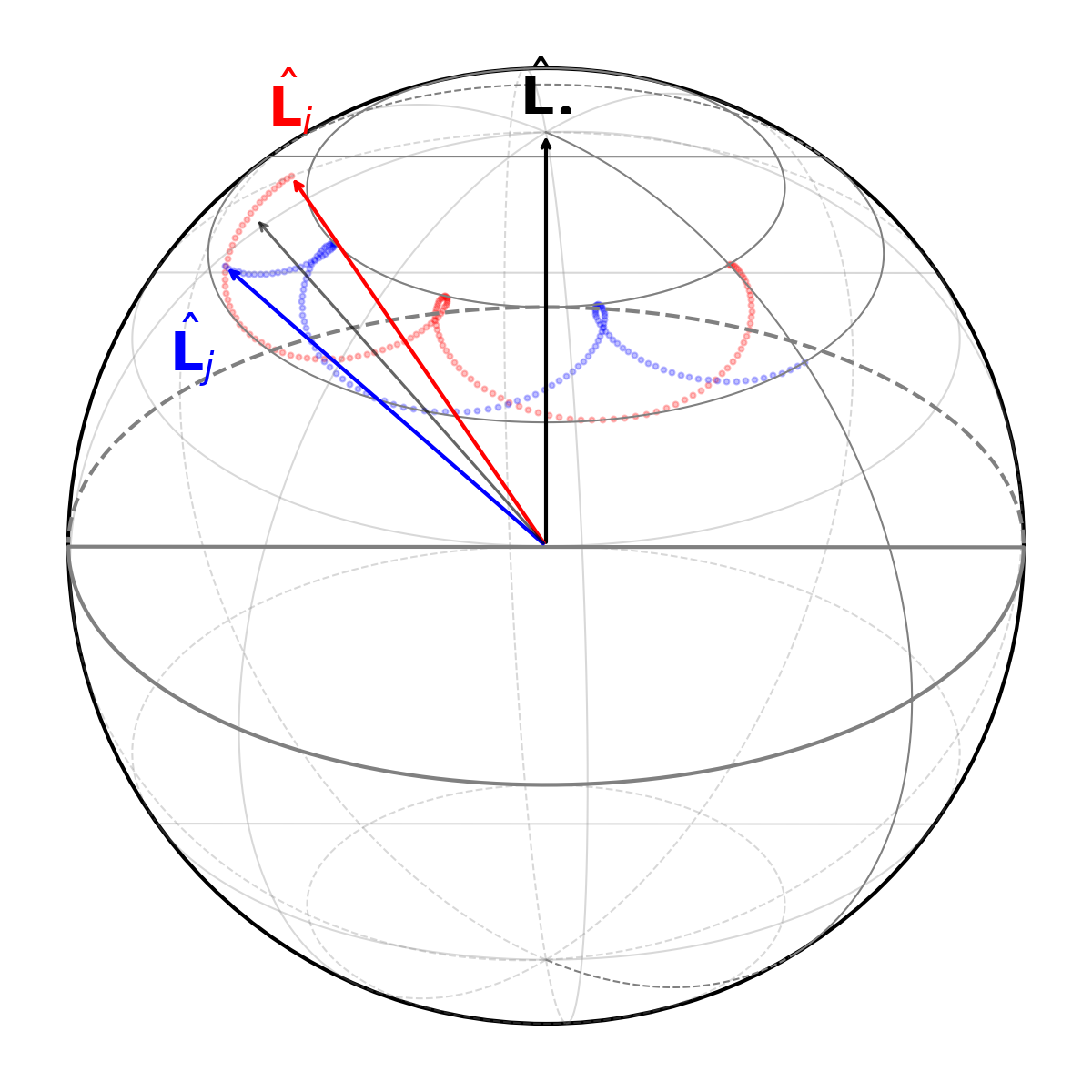}
    \caption{Coupled angular momentum pair}
    \label{fig:coupled}
  \end{subfigure}\hfill
  \begin{subfigure}[b]{0.46\linewidth}
    \centering
    \includegraphics[width=\linewidth]{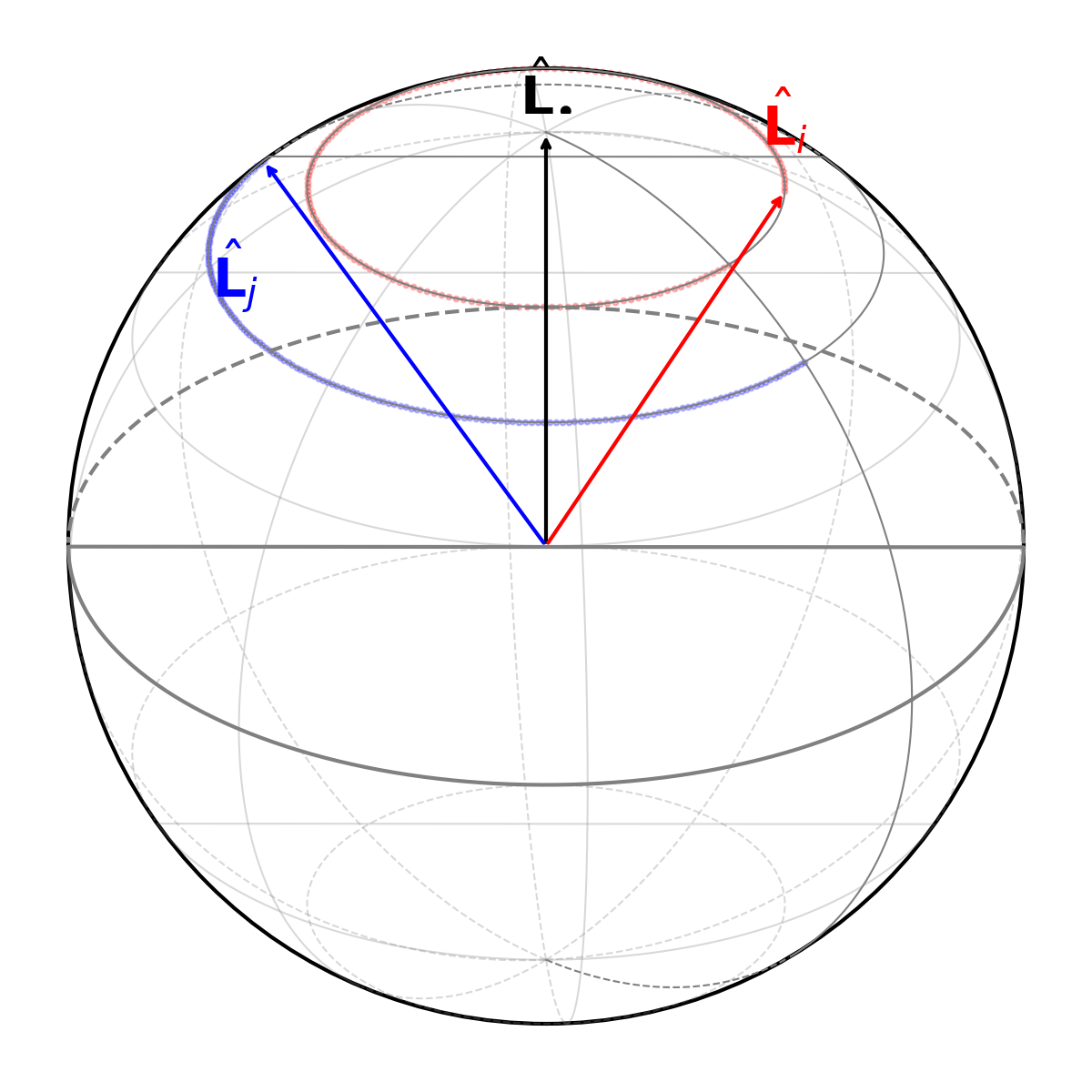}
    \caption{Decoupled pair}
    \label{fig:decoupled}
  \end{subfigure}

  \caption{Angular-momentum vectors on the unit sphere.}
  \label{fig:sphere_demo}
\end{figure*}

We consider a system of $4$ masses: the SMBH, two similar masses with indices $i$ and $j$ and one massive perturber, whose mass and angular momentum are $m_\IMBH$ and $\L_\IMBH$. We keep the hierarchy $m_i,m_j \ll m_\IMBH \ll \MSMBH$. The initial configuration of the system and its possible evolutionary outcomes are illustrated in Fig.~\ref{fig:sphere_demo}. Under the assumption where the angular momentum of the massive perturber dominates the total angular momentum of the system, $|\L_\IMBH| \gg |\L_i+\L_j|$ for all $i$,
equation \eqref{eq:EOM} implies that the total angular-momentum vector of any $i$-$j$ pair will precess around the angular-momentum vector of the perturber with angular velocity $\Omega_{\IMBH}$. Let us see what will happen to the relative difference $\L_i - \L_j$.

In the restricted three-body problem, a binary may be separated by a massive perturber if it crosses within the Hill radius, or tidal radius, defined to be the distance where the tidal force of the perturber equals the force exerted by the binary companion \citep{Hill1887,Hills1988,Arnoldetal2006}. Here, by analogy, we may compare the magnitude of the torques or the angular velocities. For example, if the frequency of rotation of $\L_i$ and $\L_j$ about each other, $\Omega_{i,j}$, is much faster than $\Omega_{i,\IMBH}$ and $\Omega_{j,\IMBH}$, then it is clear that they will rotate about each other, while their sum $\L_i+\L_j$ will precess about $\L_\IMBH$, so the angular momentum binary should persevere (Fig.~\ref{fig:coupled}). Conversely, if $\Omega_{i,\IMBH}$ and $\Omega_{j,\IMBH}$ dominates, then the components of the angular momentum binary will rotate about $\L_\IMBH$ individually with their respective angular frequencies and the angular momentum binary would be disrupted (Fig.~\ref{fig:coupled}). This, again, is in direct analogy to the standard Hill criterion of time-scale comparison, between the orbital period of the ``outer'' orbit and that of the ``inner'' orbit \citep{Hill1887,Hills1988}. 
Thus, in order to find the condition for disruption, we compare magnitudes of the torque due to the pair with the relative torque: we require
\begin{equation}\label{eqn:fundamental criterion}
\left|\mathbf{\Omega}_{i,j}\times\L_i\right| \leq \left| \left(\mathbf{\Omega}_{i,\IMBH}\times\L_i - \mathbf{\Omega}_{j,\IMBH}\times\L_j \right) \cdot \Delta\hat{\L}\right|
\end{equation}
for disruption,
where the scalar product with $\Delta \hat{\L}=\Delta \L/|\Delta \L|$ describing the direction of $\Delta \L \equiv \L_i - \L_j$ arises because only the component parallel to $\Delta \L$ can lead to an instability -- the perpendicular components can never change the magnitude of $\Delta \L$. 

Generally, the angular frequency $\Omega$ can depend on multiple variables, the relevant ones being the orbital inclination $\theta$, semi-major axis $a$ and the eccentricity $e$. For simplicity we consider the equal-mass case for now, but generalise below. The coupling coefficients $\mathcal{J}_{ij\ell}$ are maximised for nearly equal semi-major axes, nearly co-planar and nearly circular orbits. So, for the pair to survive the torques from the perturber, $a_i$ and $a_j$, $e_i$ and $e_j$, and $\theta_i$ and $\theta_j$ must all be close to each other, whence
\begin{align}
\left|\mathbf{\Omega}_{i,j}\times\L_i\right| \leq& \left| \left[\mathbf{\Omega}_{\IMBH}\left( \theta_{\IMBH} + \frac{\theta_b}{2}, a + \frac{\Delta a}{2}, e + \frac{\Delta e}{2} \right) \times \L_i \right.\right.\nonumber\\ &-\left.\left. \mathbf{\Omega}_{\IMBH}\left( \theta_{\IMBH} - \frac{\theta_b}{2}, a - \frac{\Delta a}{2}, e - \frac{\Delta e}{2} \right) \times \L_j \right]\cdot \Delta\hat{\L}\right|.
\end{align}
For convenience, we introduced the average semi-major axis $a$ and eccentricity $e$ of the pair, such that $\Delta a = a_j - a_i$, $\Delta e = |e_j - e_i|$ (recall that we are concerned with the equal-mass case here).
Here, $\theta_b$ is the angle between the components of the pair, $\theta_\IMBH$ is the angle between total angular-momentum vector of the pair and the perturber; we further take the orientation of the relative angular momentum vector to be along a meridian of the unit sphere. We denote by index $i$ the component of the pair with smaller semi-major axis, i.e.~$a_i \leq a_j$. Expanding this in all variables to the first order\footnote{In the overlapping case that $\ell^2{J}_{ij\ell}$ have narrow spike-like singularities for identical orbits as a function of $\Delta a$ and $\Delta e$, which clearly cannot be Taylor-expanded around the identical orbits with the IMBH, however this applies to a narrow range in parameter space. Additionally, we restrict ourselves to the the case $\Delta a = \Delta e = 0$ below in this study.} turns inequality \eqref{eqn:fundamental criterion} into
\begin{align}
\label{eq:torques_rel_taylor}\left|\mathbf{\Omega}_{i,j}\times\L_i\right| \leq& \Bigg|\Bigg[ \mathbf{\Omega}_{\IMBH}\left( \theta_{\IMBH}, a, e\right) \times (\L_i - \L_j) 
\\
-& \frac{1}{2} \left( \frac{\partial \mathbf{\Omega}_{\IMBH}}{\partial \theta} \theta_b + \frac{\partial \mathbf{\Omega}_{\IMBH}}{\partial a} \Delta a + \frac{\partial \mathbf{\Omega}_{\IMBH}}{\partial e} \Delta e \right) \times (\L_i + \L_j) \Bigg]\cdot \Delta\hat{\L}\Bigg|.\nonumber
\end{align}
The first term on the right drops out, because it is always perpendicular to $\Delta \L$ -- it can only rotate $\Delta \L$. We finally find
\begin{align}
\Omega_{i,j}\left(\theta_b, \frac{a_i}{a_j}\right)L_i\sin\theta_b \leq&  \frac{1}{2} \left( \frac{\partial \Omega_{\IMBH}}{\partial \theta} \theta_b\kappa + \frac{\partial \Omega_{\IMBH}}{\partial a} \Delta a + \frac{\partial \Omega_{\IMBH}}{\partial e} \Delta e \right)
\nonumber\\&\times |\L_i+\L_j| \sin\theta_\IMBH,
\label{eq:torques_mag}
\end{align}
where $\kappa\leq1$ is a geometric factor of order unity which accounts for the projection of the difference vector $\Delta \hat{\L}$ onto the gradient of the precession frequency. Since the precession frequency $\Omega_\IMBH$ depends on the inclination $\theta$ relative to $\L_\IMBH$, the differential torque is maximized when $\Delta \hat{\L}$ is aligned with the direction of changing $\theta$ (i.e., along a meridian of the sphere defined by the pole $\L_\IMBH$; see Fig.~\ref{fig:sphere_demo}). Conversely, the effect vanishes if $\Delta \hat{\L}$ is azimuthal (along a latitude). $\kappa$ captures the time-averaged orientation of the precessing pair $(\L_i, \L_j)$ with respect to this frame; its effective value can be calibrated using N-ring simulations. The geometry of the system gives constrains on the maximum relative inclination of the pair -- when the system remains coupled, the relative angle is constant; but when it is decoupled, the relative angle varies between $\cos(\theta_b)=\cos(\max[\theta_{1}\pm\theta_{2}])$ where ``max'' stands for the configuration where the initial polar angles $\theta_{1,2}$ are aligned according to a meridian as in Figure~\ref{fig:set-up}.

Thus, given the formul{\ae} for the angular frequencies, one can use inequality \eqref{eq:torques_mag} to determine whether or not the pair would be disrupted. Analytical expressions can be obtained for $\mathbf{\Omega}_{i,j}$ when the orbits have large radial separations (dominated by the $\ell=2$ harmonic) and for small radial separations (dominated by high-order multipole moments $\ell\to \infty$). The mutual torque between two particles is the strongest when $a_i\to a_j$, so we will only consider the asymptotic limit for the left-hand side of Eq.~\eqref{eq:torques_mag}.

We now restrict our attention to a special -- but analytically tractable -- configuration, in which the two bodies share the same semi-major axis and eccentricity, \emph{viz.} \(a_i=a_j=a\) and \(e_i=e_j=e\).  
In such a synchronous pair, disruption can arise only from changes in their mutual inclination.  
The disruption criterion therefore reduces to
\begin{equation}
\Omega_{i,j}\!\left(\theta_b,1\right)m_i\le\;\frac{1}{2}
\kappa\,\left|\sin\theta_{\IMBH}\,
\frac{\partial \Omega_{\IMBH}\!\left(\theta_{\IMBH},a/a_{\IMBH}\right)}{\partial \theta}\right|(m_i+m_j),
\label{eq:synchronous}
\end{equation}
where we have used the small-angle approximation \(\sin\theta_b\simeq\theta_b\) and $|\L_i+\L_j| \simeq L_i + L_j$.  
Inequality~\eqref{eq:synchronous} can be re-arranged to solve either (i) for \(\theta_b\), yielding the minimum relative inclination at which the pair decouples for a given distance from the perturber, or (ii) for \(a/a_{\IMBH}\), giving the critical separation where disruption occurs, for a specified value of \(\theta_b\).

When the differential torque due to the IMBH significantly exceeds the mutual torque of the pair, the disruption time-scale can be estimated as
\begin{equation}
t_{\rm dis}(a,\theta_{\IMBH})
\simeq
\frac{\pi}{|\Omega_{i,\IMBH} - \Omega_{j,\IMBH}|}
\;\approx\;
\frac{\pi}{\theta_b\,\bigl|\partial \Omega_{\IMBH}\!\left(\theta_{\IMBH},a/a_{\IMBH}\right)/\partial \theta\bigr|},
\label{eq:t-disr}
\end{equation}
which corresponds to the time required for the relative precession phase around the IMBH to drift by $\Delta\phi\simeq\pi$, i.e. for the pair to reach its maximum angular separation on the sphere.

\subsection{Synchronous circular orbits}
\label{subsec:circular}

The circular case deserves special interest as it can be solved analytically for both the asymptotic limit and for the quadrupole interactions, as we demonstrate in Appendix~\ref{app:energy}. Here we denote $a_i=a_j=r$ and $a_\IMBH = r_\IMBH$. 

As we show in Appendix~\ref{app:energy}, the angular frequency between the pair (for small relative angles $\theta_b \ll 1$), is given by
\begin{equation}
\Omega_{i,j} = 2\omega_i \frac{m_{j}}{\MSMBH} \frac{1}{\theta_b^2}.
\label{eq:omega_circ_asymp}
\end{equation}

Substituting the Eq.~\eqref{eq:omega_circ_asymp} for the left-hand-side of the Eq.~\eqref{eq:synchronous} the quadrupole approximation in Eq.~\eqref{eq:quad_circ} for the right-hand into the
disruption condition~\eqref{eq:synchronous} and retaining the small-angle
approximation \(\sin\theta_b\simeq\theta_b\), we determine the conditions under which the pair decouples due to the perturber’s influence. As remarked above, the inequality~\eqref{eq:synchronous} can be re-arranged to solve for the relative angle \(\theta_b\) and the distance ratio \(\r/r_{\IMBH}\).
First of all, solving for the relative angle \(\theta_b\), we obtain:
\begin{equation}
\theta_b \ge
\begin{cases}
\displaystyle
\left(\frac{4}{3\pi\kappa}\right)^{1/2}
\left(\frac{2m_im_j}{m_{\IMBH}(m_i+m_j)}\right)^{1/2}
\left(\frac{r}{r_{\IMBH}}\right)^{-3/2}
\frac{1}{\lvert\sin\theta_{\IMBH}\rvert},
& r<r_{\IMBH},\\[1.2em]
\displaystyle
\left(\frac{4}{3\pi\kappa}\right)^{1/2}
\left(\frac{2m_im_j}{m_{\IMBH}(m_i+m_j)}\right)^{1/2}
\left(\frac{r}{r_{\IMBH}}\right)
\frac{1}{\lvert\sin\theta_{\IMBH}\rvert},
& r\ge r_{\IMBH}.
\end{cases}
\label{eq:thetab}
\end{equation}
Secondly, solving for the distance ratio ${r}/{r_{\IMBH}}$, we find:
\begin{equation}
\begin{cases}
\displaystyle
\frac{r}{r_{\IMBH}} \ge
\left(\frac{4}{3\pi\kappa}\right)^{1/3}
\left(\frac{2m_im_j}{m_{\IMBH}(m_i+m_j)}\right)^{1/3}
\theta_b^{-2/3}
\lvert\sin\theta_{\IMBH}\rvert^{-2/3},
& r<r_{\IMBH},\\[1.2em]
\displaystyle
\frac{r}{r_{\IMBH}} \le
\left(\frac{3\pi\kappa}{4}\right)^{1/2}
\left(\frac{2m_im_j}{m_{\IMBH}(m_i+m_j)}\right)^{-1/2}
\theta_b
\lvert\sin\theta_{\IMBH}\rvert,
& r\ge r_{\IMBH}.
\end{cases}
\label{eq:radius_ratio}
\end{equation}

This expression provides the critical distance-ratio at which the pair decouples due to the perturber’s torque, given a specific relative angle \(\theta_b\). The expressions above are good approximations for $r/r_{\IMBH} \leq 0.3$, otherwise the quadrupole approximation for $\boldsymbol{\Omega}_\IMBH$ fails (cf.~Appendix \ref{app:energy}). One can obtain similar conditions for larger values of the distance ratios using the asymptotic expansions with a correction for the low-order terms (equation \eqref{eq:S'_ij}), which can be solved analytically for the relative angle and numerically for the distance ratio (see Appendix~\ref{app:energy}).

Observe that in the quadrupole approximation the critical inclination scales as 
$\theta_b\propto 1/|\sin\theta_\IMBH|$, so pairs whose orbital planes are
nearly \emph{perpendicular} to the IMBH are the most likely to be disrupted. The picture changes for larger distance ratios when higher-order harmonics become important. 

To test inequality~\eqref{eq:synchronous} we ran a suite of \nring{} 
integrations\,(described in Appendix~\ref{app:n-ring}) tailored to the Milky-Way nucleus
with $\MSMBH=4\times10^{6}\,M_\odot$. Two experiment sets were
performed: (i)~A \emph{quadrupole} set with $r/r_\IMBH=1/3$ and
$m_\IMBH=5\times10^{4}\,M_\odot$;  
(ii)~a \emph{multipole} set with
$r/r_\IMBH=2/3$ and $m_\IMBH=2\times10^{3}\,M_\odot$ -- a regime in which terms
with $\ell>2$ are important.  
Both adopt $a_\IMBH=0.15$\,pc and $m_i=m_j=10\msun$. For each set we
generated $\simeq 10^{4}$ three-ring realisations where the angles between the pair's angular momenta with respect to the perturber were drawn
from a distribution uniform in $\cos\theta_\IMBH$.

\begin{figure*}
  \centering
  \begin{subfigure}[t]{0.95\linewidth}
    \centering
    \includegraphics[width=\linewidth]{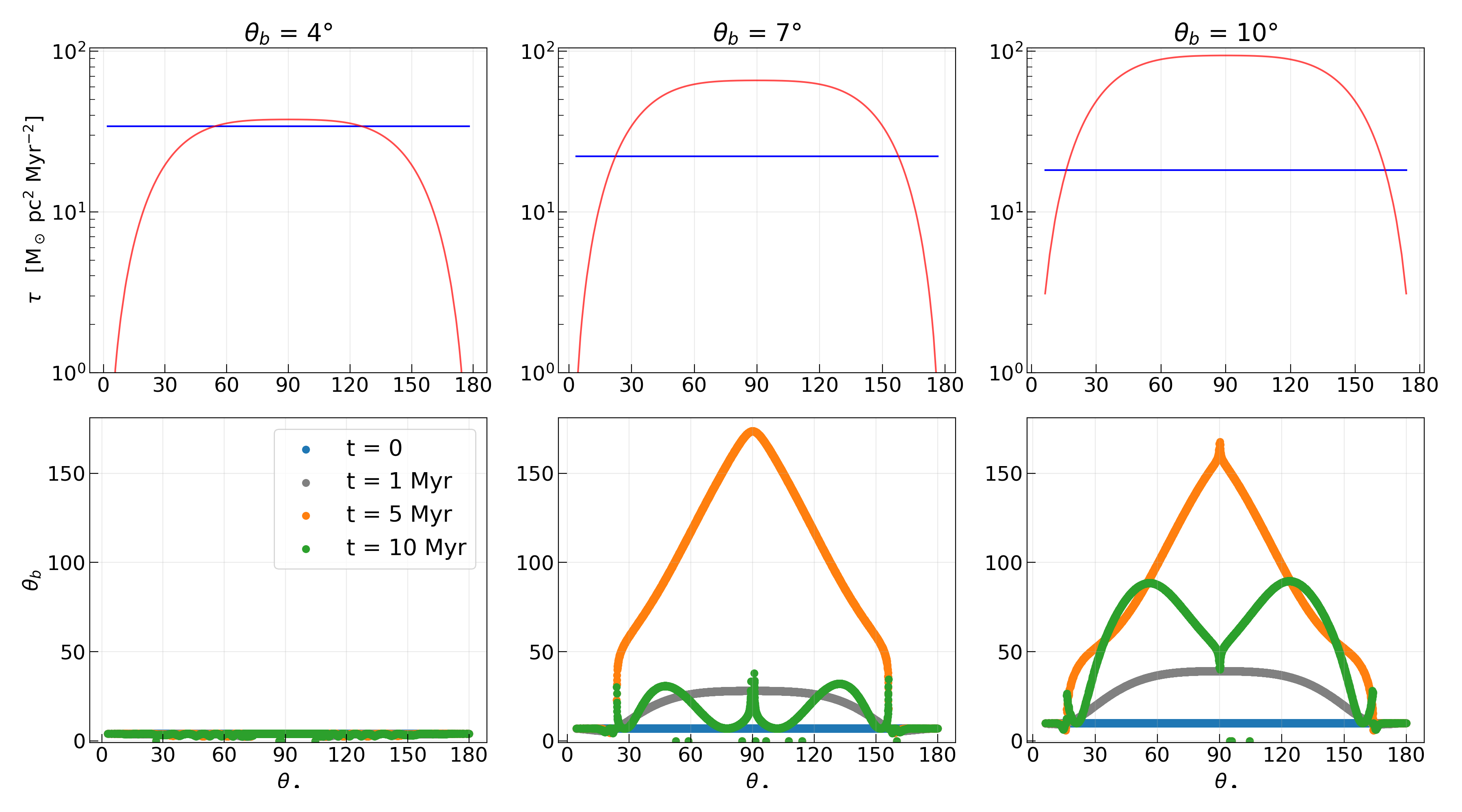}
    \caption{$a/a_{\IMBH}=1/3$,  $a_{\IMBH}=0.15$ pc, $m_{\IMBH}=5\times10^{4}\,M_{\odot}$; $\Omega_\IMBH^{-1}(\pi/4,a/a_\IMBH)\approx3.5\times10^{5}$~yr.}
    \label{fig:torques_quad}
  \end{subfigure}

  \vspace{1em}

  \begin{subfigure}[t]{0.95\linewidth}
    \centering
    \includegraphics[width=\linewidth]{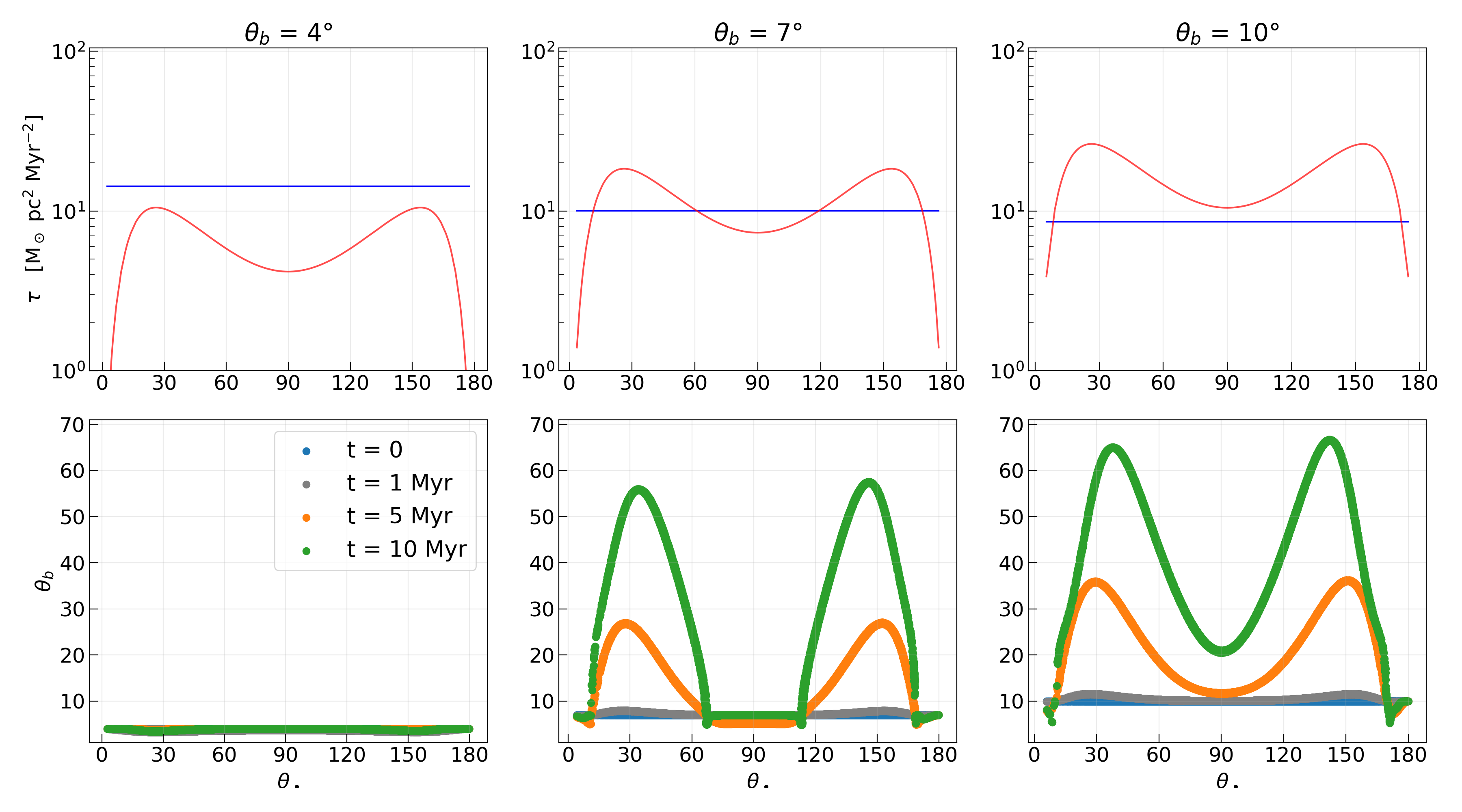}
    \caption{$a/a_{\IMBH}=2/3$, $a_{\IMBH}=0.15$ pc, $m_{\IMBH}=2\times10^{3}\,M_{\odot}$;  $\Omega_\IMBH^{-1}(\pi/4,a/a_\IMBH)\approx20$~Myr.}
    \label{fig:torques}
  \end{subfigure}

  \caption{%
    A comparison of analytical prediction by Equation~\eqref{eq:synchronous} with the results of \nring simulations. \emph{Top panels}: magnitudes of the pair's internal torque (blue) and of the relative torque from the IMBH (\(\theta_\IMBH\); red) as functions of the pair’s inclination with respect to the IMBH, for mutual inclinations of \(\theta_b=4^\circ\), \(7^\circ\), and \(10^\circ\), computed using Equation~\eqref{eq:synchronous} for circular orbits.  
    \emph{Bottom panels}: the mutual inclination of every pair versus \(\theta_\IMBH\) at \(t=0\) (blue), 1~Myr (grey), 5~Myr (orange), and 10~Myr (green) obtained from \nring integrations of the same configurations.  
    The simulations confirm the analytic criterion: all pairs remain bound for \(\theta_b=4^\circ\), whereas those entering the unstable (red-above-blue) region are disrupted for \(\theta_b=7^\circ\) and \(10^\circ\).
In the quadrupole-dominated regime (panel a) the torque difference exhibits a single clear maximum around $\theta_{\IMBH}=90^\circ$, whereas in the general case (panel b) higher-order multipoles introduce second prominent peaks. In both configurations the analytic disruption criterion (red curve exceeding blue) accurately predicts which angular momentum binaries become unbound by 10\,Myr.}
\label{fig:rel_torques}
\end{figure*}

Figure~\ref{fig:rel_torques} confirms the validity of
condition~\eqref{eq:synchronous}. %
In the top rows of both panels (a) and (b), the blue curves show the torques within each pair (the left‑hand side of Eq.~\eqref{eq:synchronous} multiplied by $|\bm{L_i}|/m_i$), 
while the red curves show the differential torques exerted by the perturber (the right‑hand side of Eq.~\eqref{eq:synchronous} multiplied by $|\bm{L_i}|/m_i$). 
Each column corresponds to an initial mutual inclination between the two objects of $4^{\circ}$, $7^{\circ}$, and $10^{\circ}$, respectively for $\kappa=0.5$.
According to condition~\eqref{eq:synchronous}, a pair becomes unbound where the red curve lies above the blue one.
The bottom rows of both panels present \nring simulations for the same initial configurations. 
Each pair was placed uniformly on the unit sphere with initial mutual inclinations of $4^{\circ}$, $7^{\circ}$, and $10^{\circ}$, and the three rings were evolved according to the equations of motion \eqref{eq:EOM}. 
The panels display the mutual inclination of each pair versus its initial inclination to the perturber at four epochs: $t=0$, 1, 5, and $10\,\mathrm{Myr}$. 
The numerical results align beautifully with the analytic prediction: all pairs remain bound in the $4^{\circ}$ case (left column)\footnote{We note that here, the analytic curve for the differential torque marginally exceeds the internal torque near $\theta_\IMBH=90^\circ$, which would formally imply instability. The fact that the simulations show these pairs remaining bound indicates that the effective geometric factor is slightly smaller than the adopted value of $\kappa=0.5$.}, whereas those located above the blue–red intersection in the $7^{\circ}$ and $10^{\circ}$ cases become unbound.

Fig.~\ref{fig:torques_quad} illustrates the \emph{quadrupole‐only} prediction obtained by retaining solely the \(\ell=2\) term and adopting a radius ratio \(r/r_{\IMBH}=a/a_{\IMBH}=1/3\). We clearly see one peak for the differential torque at $\theta_\IMBH=90^\circ$. By contrast, Fig.~\ref{fig:torques} displays the outcome when the multipole expansion is summed to \(\ell_{\max}=200\) at the larger separation \(r/r_{\IMBH}=a/a_{\IMBH}=2/3\).  The additional harmonics introduce the secondary peaks visible in the red perturber-curve, yet in both panels the crossing of the blue (pair) and red (perturber) curves still marks the critical boundary, beyond which the \nring simulations in the lower rows show the binaries becoming unbound.

Additionally, Fig.~\ref{fig:torques_quad} illustrates a case with relatively fast nodal precession $\Omega_\IMBH^{-1}(\pi/4,a/a_\IMBH)\approx3.5\times10^{5}$~yr about the angular momentum of the perturber with mass $m_{\IMBH}=5\times10^{4}\,M_{\odot}$. The figure shows that for the initial pair's inclination of 7$^\circ$, the disrupted pairs reached maximum separation at $t = 5$~Myr (orange curve in the second row), while at 10~Myr (green curve) the angular separations are low again, illustrating the oscillatory behaviour for the disrupted pairs between $\theta_b$ and $\theta_1+\theta_2$ ($2\pi - (\theta_1+\theta_2)$ when $\theta_1+\theta_2>180^\circ$), as evident from the geometry of the problem (see Fig.~\ref{fig:sphere_demo}). A similar pattern is observed for the initial angular separation of $10^\circ$, but with a slightly faster disruption time-scale. On the other hand, Fig.~\ref{fig:torques} depicts the case where nodal precession is much slower,
$\Omega_\IMBH^{-1}(\pi/4,a/a_\IMBH)\simeq20$\,Myr for a perturber with mass $m_{\IMBH}=2\times10^{3}\,M_{\odot}$. 
The orange and green curves show that pairs with low‐inclination
orbits have already reached their maximum separation ($\cos\theta_b\lesssim\cos(\theta_1+\theta_2)$), whereas those that
start at higher inclinations are still progressing toward that point ($\cos\theta_b\ll\cos(\theta_1+\theta_2)$), especially the ones closer to the equatorial plane.

\begin{figure*}
    \centering
    \includegraphics[width=\linewidth]{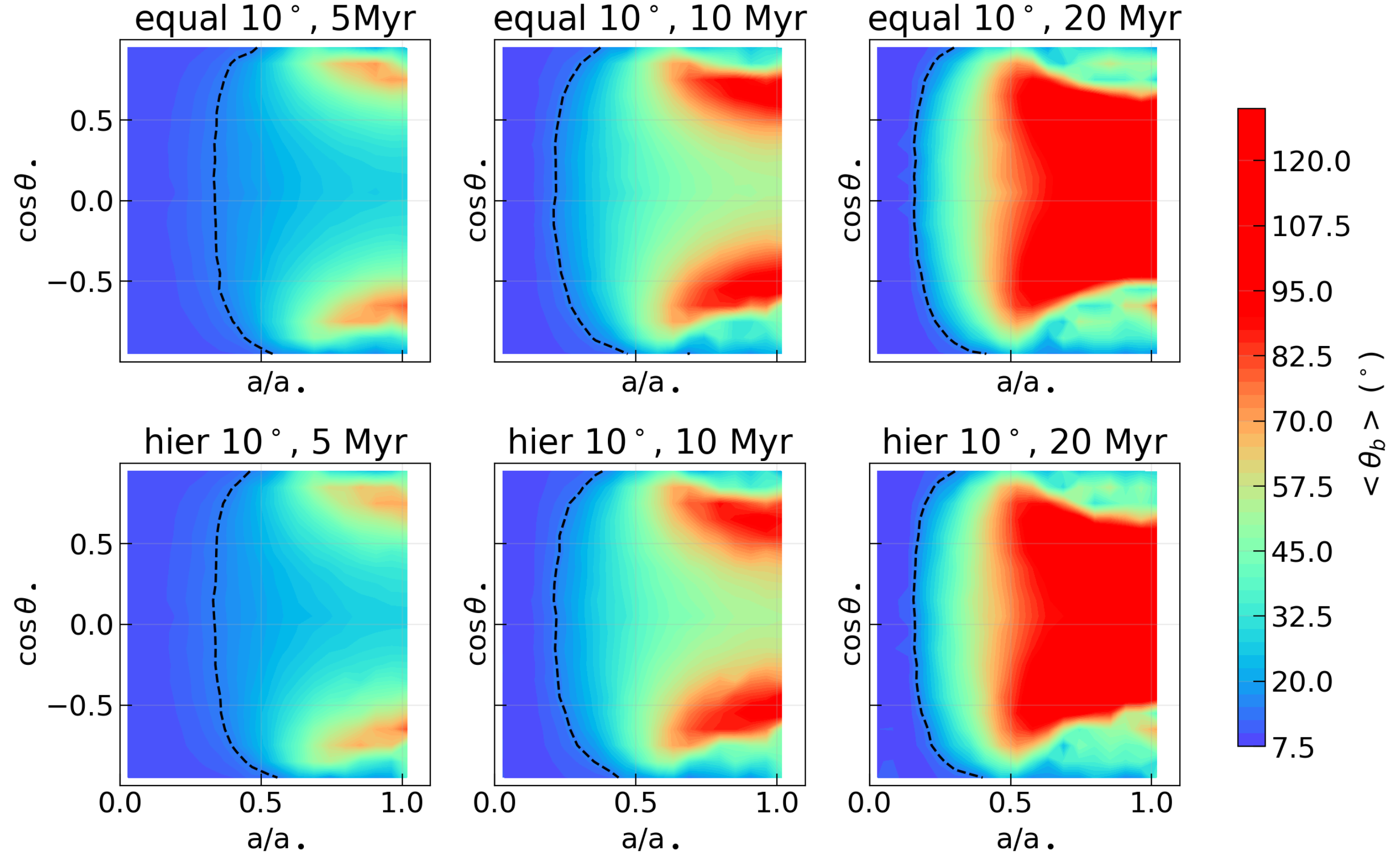}
    \caption{Average mutual inclination angle between pairs as a function of cosine of inclination with respect to the perturber and distance ratio to the perturber for initial mutual inclination of $10^\circ$. \emph{Top panels}: the equal mass case, \emph{bottom panels}: hierarchical case ($m_i\gg m_j$). Each column corresponds to 5,10 and 20 Myr of the evolution. The dashed lines on each of the panels show the region where the average inclination of the pair does not exceed $15^\circ$. The plot shows results for \texttt{N}-Ring simulations with an IMBH mass of $m_\IMBH=10^5\msun$ at a distance $r_\IMBH=1.5$ pc from the SMBH of mass $4\times10^6\
    \msun$.}
    \label{fig:angle_avg}
\end{figure*}

\subsection{Hierarchical mass-ratio pairs}
Let us consider the case where $a_i=a_j$ as previously, but with $m_i \gg m_j$. 
In this case, $\L_j$ precesses about $\L_i+\L_j \approx \L_i$ in the absence of an IMBH, while conserving $\hat{\L}_i \cdot \hat{\L}_j$; this is thus a nodal precession about $\L_i$. The analogy with tides (equivalently, the restricted three-body problem) implies that equation \eqref{eqn:fundamental criterion} should reduce to a comparison between the torque $\boldsymbol{\Omega}_{ij} \times \L_j$ and the torque-difference $\Omega_{j,\IMBH} \times \L_j$ between opposite positions in the orbit of $\L_j$ about $\L_i$ -- when the argument of the ascending node of particle $j$ is $0$ and $\pi$. We denote the difference between the corresponding values of $\L_j$ by $\delta \L_j$, which is necessarily perpendicular to $\L_i$. 
The above reasoning still applies, and therefore the stability condition for this pair is still 
\begin{equation}\label{eqn:vector product}
    \abs{\sum_{\ell=2}^{\infty}\mathcal{J}_{ij\ell}P_{\ell}'\left(\cos \theta_b\right)\sin\theta_b} \leq 
    \abs{\sum_{n\in\{x,y,z\}} \delta L_j^n\frac{\partial \boldsymbol{\Omega}_\IMBH}{\partial L_j^n} \times \L_j },  
\end{equation}
where $L_j^n$ is the $n$-th Cartesian component of $\L_j$ (with $n\in\{x,y,z\}$), and we have dropped the $\Omega_{j,\IMBH} \times \delta\L_j$ term, whose scalar product with $\delta \L_j$ vanishes. The angular-momentum difference $\delta \L_j$ is related to $\theta_b$ via $\delta L_j = 2L_j \sin \theta_b$. 

We now simulate a system containing an IMBH, of mass
$m_\IMBH=10^{5}\,M_\odot$, located at $r_\IMBH=1.5$\,pc.  
Stellar pairs are initialised with semi-major axes distributed
between 0.04 and 1.5\,pc.  Two data sets are examined:
\begin{enumerate}
\item \textbf{Equal-mass case:} $m_i=m_j=10\,M_\odot$, identical to the
      masses used in the previous models.
\item \textbf{Hierarchical case:} $m_i=20\,M_\odot$ and
      $m_j=0.01\,M_\odot$.  This configuration preserves the total pair
      mass but allows the lighter component to behave effectively as a
      test particle.
\end{enumerate}
Both data-sets were drawn from the same distributions and the pairs start from an initial angular separation of 10$^\circ$. Again, we run $\simeq10^4$ individual simulations for each case and stack results together for the analysis.

In Fig.\,\ref{fig:angle_avg} we plot the mean mutual inclination
\(\langle\theta_b\rangle\) in the \((\cos\theta_\IMBH,\;a/a_\IMBH)\) plane at
\(t = 5,\;10,\) and \(20\) Myr (three columns).  The top row shows the
\emph{equal-mass} model, the bottom row -- the \emph{hierarchical} model; as
expected, the two are nearly indistinguishable.  
For each panel the dashed contour marks \(\langle\theta_b\rangle =
15^{\circ}\); regions to left of this curve remain effectively bound.  
The colour scale traces the growth of the pair separation with time and
highlights the migration of the differential-torque maxima: moving from
larger to smaller \(a/a_\IMBH\) (right to left), the two peaks drift toward
\(\theta_\IMBH = 0^{\circ}\) and \(180^{\circ}\).  
At the smallest radii the dashed contour is single-peaked, signalling
the transition to the quadrupole-dominated regime.

\subsection{Elliptical orbits}
\label{subsec:elliptical orbits}

\begin{figure*} 
    \centering
    \includegraphics[width=0.99\linewidth]{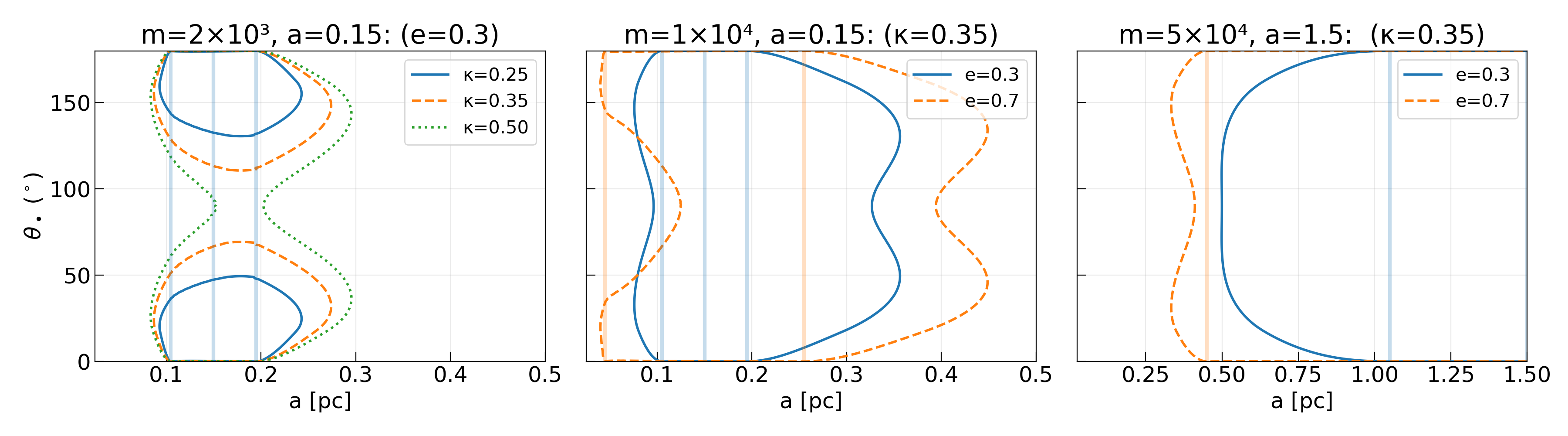}
    \caption{Contour lines show the analytical boundary of stability for a pair of circular orbits against disruption by an eccentric IMBH for initial mutual inclination of $7^\circ$, i.e. regions where the torque between a pair of bodies equals the right hand side of the Eq.~\eqref{eq:synchronous} for eccentric orbits, calculated from the analytical expressions in section \ref{subsec:elliptical orbits}. Vertical lines show the pericentre, the semi-major axis, and the apocentre of the IMBH.  The IMBH mass, eccentricity, and $\kappa$ parameter are indicated in the panels.}
    \label{fig:torques_analytic}
\end{figure*}

In general the orbits can be divided into two basic
geometrical classes: radially non-overlapping ($r_\mathrm{a} < r_{\rm p,\IMBH}$ or $r_\mathrm{p} > r_{\rm a,\IMBH}$) and radially overlapping orbits ($r_{\rm a} \geq r_{\rm p,\IMBH} \text{~and~} r_{\rm p} \leq r_{\rm a,\IMBH}$) with respect to the perturber. 
The mutual torques are strongest for overlapping orbits (see \citealt{Kocsis2015}), whence, as for the circular case, we always consider only overlapping orbits for the pair, and use the asymptotic expressions to describe mutual torque within the pair. Thus, the angular frequency can be approximated as:

\begin{equation}
    \mathbf{\Omega}_{i,j}  \approx -  \frac{1}{2}k_{ij} \omega_{\rm orb,i}\cot \theta\,\Ln_j \,.
\label{eq:omega_asymp_over2}
\end{equation}
where
\begin{equation}     
    k_{ij}= \dfrac{4}{\pi^3} \dfrac{m_{j}}{\MSMBH} \dfrac{I_2}{a_j(1-e_i^2)^{1/2}}.
\end{equation}

\subsubsection{Non-overlapping orbits with respect to the perturber}

Here we examine two cases for the stellar pair: eccentric pair with overlapping orbits or a circular pair; and in both cases the perturber is eccentric. For a distant perturber we can use the quadrupole approximation. Using Eqs.~\eqref{eq:omega_asymp_over2} and \eqref{eq:quad_nonover}, we get:
\begin{equation}
\theta_b \;\ge\;
\begin{cases}
\displaystyle
\frac{4}{3\pi^{4}\kappa}\,
\frac{m_i m_j}{m_{\IMBH}(m_i+m_j)}\,
\frac{a_\IMBH^{3}\,I_2}{a^{4}}\,
\left(\dfrac{\chi}{\chi_\IMBH}\right)^{4}\,
\frac{1}{P_{3}(\chi)}\,
\frac{1}{\sin^{2}\theta_\IMBH},
& a < a_\IMBH \\[1.4em]
\displaystyle
\frac{4}{3\pi^{4}\kappa}\,
\frac{m_i m_j}{m_{\IMBH}(m_i+m_j)}\,
\frac{a\,I_2}{a_\IMBH^{2}}\,
\left(\dfrac{\chi_\IMBH}{\chi}\right)^{2}\,
\frac{1}{P_{3}(\chi_\IMBH)}\,
\frac{1}{\sin^{2}\theta_\IMBH},
& a > a_\IMBH.
\end{cases}
\label{eq:thetab_nonover}
\end{equation}
which can be simplified when both components of the pair are on circular orbits with the same radii:
When both components of the pair are on circular orbits with the same radii and eccentric perturber, we also use Eq.~\eqref{eq:quad_nonover} for the perturber, but Eq.~\eqref{eq:omega_circ_asymp_app} for the pair to get\footnote{Note that the asymptotic expressions used for the torques between pairs behave differently for circular-circular, circular-eccentric and eccentric-eccentric orbital configurations. See Appendix~\ref{App:Legendre} and Appendix~B5 of \citet{Kocsis2015}}:
\begin{equation}
\theta_b \;\ge\;
\begin{cases}
\displaystyle
\left(\frac{4}{3\pi\kappa}\right)^{1/2}
\left(\frac{2m_im_j}{m_{\IMBH}(m_i+m_j)}\right)^{1/2}
\left(\frac{r}{a_{\IMBH}}\right)^{-3/2}\frac{1}{(1-e^2_\IMBH)}
\frac{1}{\lvert\sin\theta_{\IMBH}\rvert},
& r < a_\IMBH \\[1.4em]
\displaystyle
\left(\frac{4}{3\pi\kappa}\right)^{1/2}
\left(\frac{2m_im_j}{m_{\IMBH}(m_i+m_j)}\right)^{1/2}
\left(\frac{a_{\IMBH}}{r}\right)\frac{(1-e^2_\IMBH)^{-1/2}}{\sqrt{P_3(\chi_\IMBH)}}
\frac{1}{\lvert\sin\theta_{\IMBH}\rvert},
& r \ge a_\IMBH.
\end{cases}
\end{equation}

Note that the expression for the external perturber differs from the circular case only by the factor $1/(1-e^2_\IMBH)$ suggesting similar behaviour of the stability condition. Similarly, 
we can solve this for the semi-major axis ratio.

\subsubsection{Overlapping orbits with the perturber}

In the case of overlapping orbits with the perturber, we can use the asymptotic expression~\eqref{eq:omega_asymp_over2} both for the pair and for the perturber, leading to:
\begin{equation}
\theta_b \ge \frac{1}{\kappa}\frac{2m_im_j}{m_{\IMBH}(m_i+m_j)}\frac{a_\IMBH}{a_i}\frac{I_2^{(i,j)}}{I_2^{(i,\IMBH)}}|\sin{\theta_\IMBH}|
\label{eq:theta_b_overlap}
\end{equation}
Note that due to complicated dependence of $I_2$ on the pericentre and apocentre distances, the expression can be explored only numerically to solve for the semi-major-axis ratio with respect to the IMBH. The dependence on the eccentricities of the pair and of the perturber is also contained implicitly in $I_2$. Note that $I_2/a_{\rm out}$ is shown in \citet{Kocsis2015}, bottom panels of Figure 1 therein, which exhibit two narrow singular spikes as a function of semi-major axis and eccentricity, when the periapsis and apoapsis of the pair and the perturber respectively coincide, and has a local minimum in between these spikes, and is monotonic in the region outside of both spikes.

It turns out that $I_2$ has a simple form for an embedded circular orbit (Appendix~\ref{app:I2}) allowing us to simplify the condition above significantly for a pair of circular orbits with the same radius $r$ embedded within an elliptical orbit of the perturber. In this case, we can use Eq.~\eqref{eq:omega_circ_asymp} for the pair, to obtain
\begin{equation}
\theta_b \ge \frac{1}{\sqrt{\kappa}}\frac{\pi}{2}\left(\frac{2m_im_j}{m_{\IMBH}(m_i+m_j)}\right)^{1/2}\frac{a_\IMBH^{1/2}\left[(r-r_\mathrm{p,\IMBH})(r_\mathrm{a,\IMBH} - r)\right]^{1/4}}{r}\sqrt{|\sin{\theta_\IMBH}|}.
\label{eq:theta_b_overlap simplified}
\end{equation}
Similarly to the previous cases, this expression can be corrected for lower-order terms (Appendix~\ref{app:EOM-corr}).

Figure~\ref{fig:torques_analytic} displays contour lines where the
differential torque exerted by the IMBH equals the mutual torque of the
pair, as prescribed by inequality~\eqref{eq:synchronous} for eccentric orbits.  
The three columns correspond to different IMBH configurations.  
In the left panel we vary the dimensionless pre-factor $\kappa$ for an IMBH with
$m_\IMBH=2\times10^{3}\,M_\odot$, $a_\IMBH=0.15$\,pc, and
$e_\IMBH=0.3$; vertical lines mark the IMBH’s pericentre, semi-major axis,
and apocentre, and the “islands’’ that appear near these radii arise from
higher multipole terms as the system transitions from overlapping to
non-overlapping orbits.  
The middle panel shows the same analysis for
$m_\IMBH=1\times10^{4}\,M_\odot$, $a_\IMBH=0.15$\,pc, and two
eccentricities ($e_\IMBH=0.3$, blue; $e_\IMBH=0.7$, orange), respectively, while the
right panel shows $m_\IMBH=5\times10^{4}\,M_\odot$,
$a_\IMBH=1.5$\,pc, and the same two eccentricities.  
For the more distant IMBH with moderate eccentricity
($e_\IMBH=0.3$, blue curve) the contour is single-peaked and
offset from pericentre, consistent with the quadrupole limit, whereas a
highly eccentric IMBH ($e_\IMBH=0.7$, orange curves in the middle and right
panels) produces double-peaked contours, indicating that higher-order
multipoles must be included in the non-overlapping regime.

\begin{figure*} 
    \centering
    \includegraphics[width=0.99\linewidth]{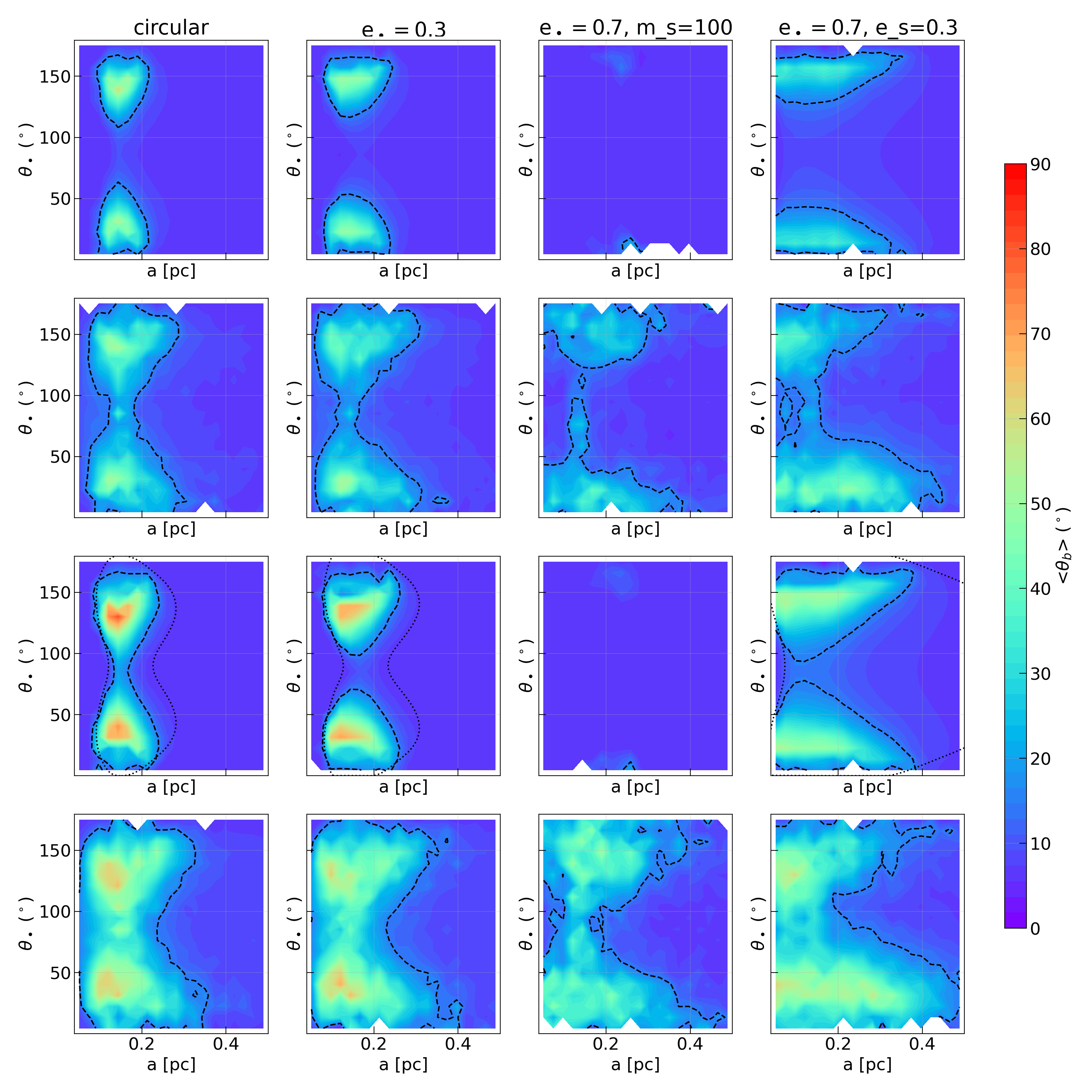}
    \caption{Average inclination angle between pairs after $5$ Myr (two \emph{top} rows) and $10$ Myr (two \emph{bottom} rows) of evolution as a function of relative distance from the IMBH and relative inclination with respect to the IMBH. In all models we used $a_\IMBH=0.15$~pc and $m_\IMBH=2\times10^3\msun$. The colour coding represents the average mutual inclination. \emph{Odd} rows -- \nring, \emph{even} rows -- \phicpu, for the identical parameters. The dashed lines show the region where $\theta_b<10^\circ$. The dotted lines in the third row show the disruption threshold computed using Eq.~\eqref{eq:synchronous}.}
    \label{fig:torques_all}
\end{figure*}

To assess the stability criterion for generally eccentric systems we carried out two
complementary types of numerical simulations: (i) orbit–averaged \nring\ integrations
(Appendix~\ref{App:simulations}) and (ii) direct $N$-body calculations with
\phicpu\ (Appendix~\ref{app:phi-cpu}).  
Figure~\ref{fig:torques_all} summarises the results.  
Rows\,1 and 3 show the \nring\ models, while rows\,2 and 4 present the
corresponding $N$-body runs; the upper two rows are snapshots at
$t=5$\,Myr, while the two lower ones are at $t=10$\,Myr.  
All simulations include an IMBH of mass
$m_\IMBH=2\times10^{3}\,M_\odot$ on a semi-major axis
$a_\IMBH=0.15$\,pc.  
The four columns represent progressively more complex
configurations:

\begin{itemize}
\item[\textit{(i)}] circular IMBH and circular stellar pairs;  
\item[\textit{(ii)}] circular pairs with an IMBH of eccentricity
      $e_\IMBH=0.3$;  
\item[\textit{(iii)}] same as (ii) but with $e_\IMBH=0.7$ and more massive
      stellar components ($m_i=m_j=100\,M_\odot$ instead of
      $10\,M_\odot$);  
\item[\textit{(iv)}] IMBH with $e_\IMBH=0.7$ and stellar pairs sharing the
      same eccentricity $e_s=0.3$.
\end{itemize}

The $N$-body integrations (rows\,2 and 4) exhibit disruption over a broader
extent of the $(\theta_\IMBH,a)$ plane and on shorter time-scales than the
\nring\ models, with irregular -- but recognisable -- boundaries that echo the
analytic contours in Fig.~\ref{fig:torques_analytic}.  
The bright “islands’’ near the IMBH pericentre, semi-major axis, and
apocentre arise from higher multipole terms, as predicted.  
Notably, in the hierarchical mass-ratio configuration (column\,iii) the
\nring\ runs show no disruption, consistent with the secular criterion,
whereas \phicpu\ reveals a sizeable unstable region.  
These discrepancies stem from resonant effects and energy–angular-momentum
exchange -- both absent in the orbit-averaged formalism -- which alter the
semi-major axes and eccentricities in the full $N$-body dynamics, and are prominent here due to the large mass of the pair.  
We quantify the regime of validity of the VRR approximation in the next
section, but emphasise that the overall morphology of the disruption
regions remains in good qualitative agreement across the analytic,
\nring, and $N$-body methods, supporting the robustness of our model.

\section{Validity of the analytical approach and VRR formalism}
\label{sec:validity}

There are two main limitations jeopardising the validity of the reasoning above: the assumption that semi-major axes are conserved might break, and even if it doesn't, the eccentricities might not be conserved with sufficient accuracy. These two assumptions stem from the hierarchy of time-scales in the system: the orbit is the fastest process allowing one to orbit-average the equations of motion, leading to a conservation of the semi-major axes (and orbital energies), and the apsidal precession time-scale is also quite fast, so one may average over that, too -- this leads to a conservation of the magnitudes of orbital angular momenta (and orbital eccentricities). 
These assumptions can fail in the presence of resonances (in the first case) or if mass-precession (or GR-precession) is not sufficiently fast (in the second case). We comment on both possibilities below.

\subsection{Three-body 1:1 resonance}
\label{subsec:resonance}
The assumption whereby the semi-major axes of the stars and the IMBH are fixed, i.e. that their orbital energies change on much longer time-scales, breaks down if one of the stars is in resonance with the IMBH. To find the regime of validity of the stability criterion for the binary, we will now estimate qualitatively how far from the resonance the semi-major axes evolve significantly. To do so, we still assume that each particle moves mostly under the influence of the SMBH, and that near the resonance, the interaction with the IMBH is the second-dominant one. 

The Hamiltonian for a star of mass $m$ on a circular orbit, moving under the gravity of the SMBH ($\MSMBH$) and the IMBH ($m_{\IMBH}$), is 
\begin{equation}
    H = -\frac{G\MSMBH m}{2a} - \frac{Gm_{\IMBH}m}{a_{\IMBH}}\frac{1}{\sqrt{1+\alpha^2-2\alpha \cos\varphi}},
\end{equation}
where $\alpha = a/a_{\IMBH}$ and $\varphi$ is the angle between the position-vectors of the star and the IMBH.\footnote{The most general case is \citep[][equations (6.2.11--6.2.12)]{Brumberg1995}
\begin{equation*}
    \begin{aligned}
        \cos \varphi & = (cc_\IMBH-ss_\IMBH)\cos(W-W_\IMBH) + c^2s_\IMBH^2 \cos(W+W_\IMBH + 2\Omega_\IMBH) 
        \\ &  
        + s^2c_\IMBH^2\cos(W+W_\IMBH - 2\Omega) + s^2s_\IMBH^2\cos(W-W_\IMBH-2\Omega + 2\Omega_\IMBH) \\ & 
        + 2csc_\IMBH s_\IMBH \left[\cos(W-W_\IMBH-\Omega) - \cos(W+W_\IMBH - \Omega - \Omega_{\IMBH})\right] 
        \\ & 
        + 2(csc_\IMBH s_\IMBH - s^2s_\IMBH^2) \cos(W-W_\IMBH),
    \end{aligned}
\end{equation*}
where $W \equiv \Omega + \omega + \lambda$ is the so-called `true anomaly in orbit' (with $\Omega$ denoting the argument of the ascending node, and $\omega$ -- the argument of pericentre), $c \equiv \cos (i/2)$ and $s \equiv \sin (i/2)$. This reduces to $\varphi = \lambda - \lambda_\IMBH$ if the orbits are co-planar.}
While we use the phase-space co-ordinates arising from the Keplerian motion of the star, this Hamiltonian is not orbit-averaged, and both $a$ and $\varphi$ change with time, generically.

\begin{figure} 
    \centering
    \includegraphics[width=0.475\textwidth]{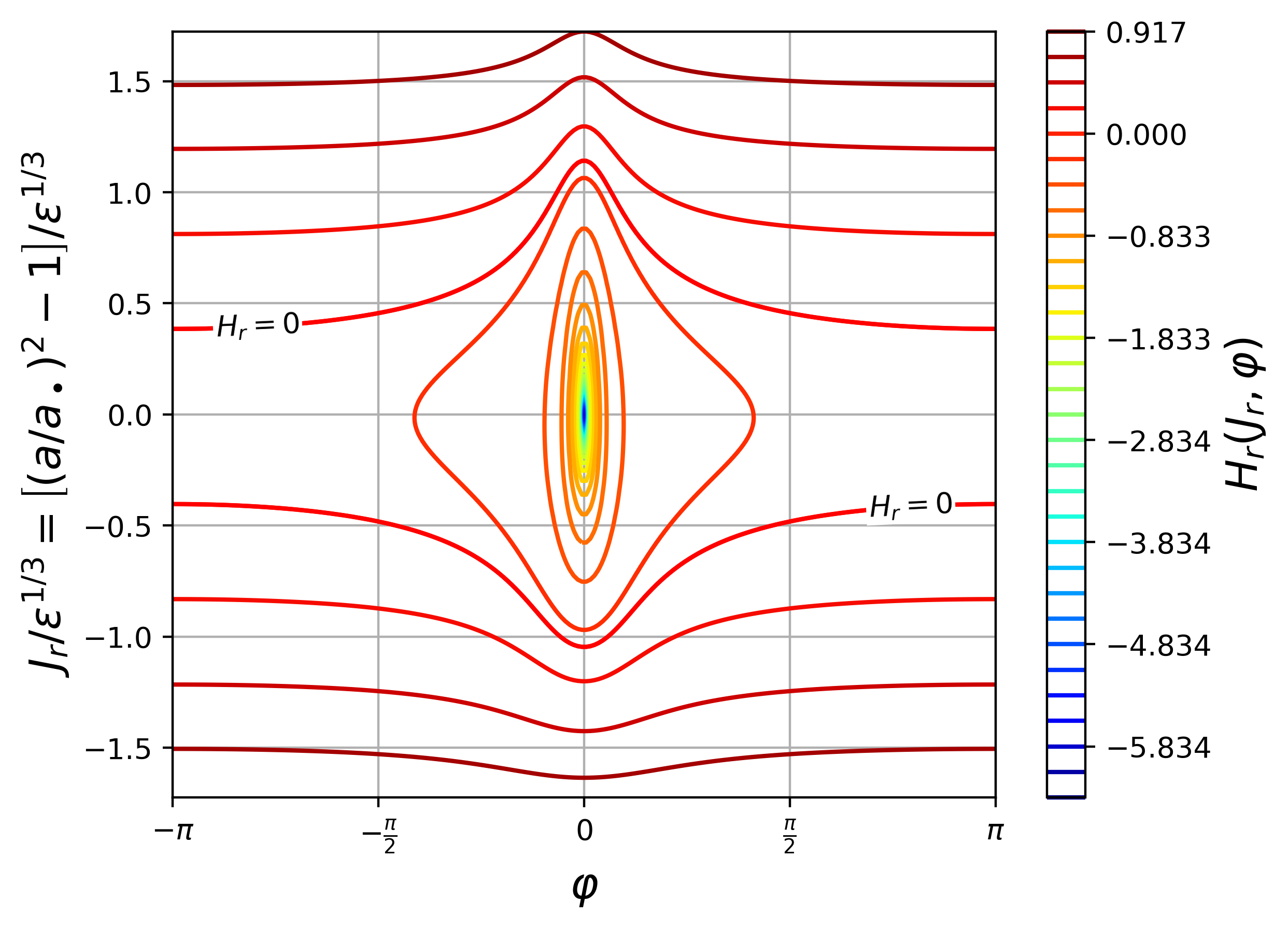}
    \caption{A level set of the Hamiltonian $H_r$ (equation \eqref{eqn:resonant Hamiltonian}), as a function of the resonant action $J_r = a^2/a_{\IMBH}^2 - 1$ and the angle $\varphi$ (for $\eps = m_{\IMBH}/\MSMBH= 0.1$). The ordinate is re-scaled by $\eps^{1/3}$, as found in equation \eqref{eqn: epsilon to 1 third}.}
    \label{fig:separatrix}
\end{figure}

Defining $\eps \equiv m_{\IMBH}/\MSMBH$ and $J^2 \equiv \alpha$, we have (up to an additive constant)
\begin{equation}
    H \mapsto \frac{G\MSMBH m}{a_{\IMBH}}\left[ - \frac{1}{2J^2} - \frac{\eps}{\sqrt{1+\alpha^2-2\alpha \cos\varphi}}\right].
\end{equation}
If we assume that the orbits are co-planar, too, then $\varphi$ is just the mean-anomaly difference of the the two orbits (up to an additive phase); otherwise, it is a trigonometric function of them and of the inclinations. The canonical momentum corresponding to $a$ is the Delaunay variable $\Lambda = m\sqrt{G\MSMBH a_\IMBH}J$, and its conjugate position variable is the mean anomaly $\lambda$; to focus on the resonance, however, we now change variables from these to $J_r \equiv J-1$ and its conjugate angle, $\lambda_r = \lambda - \lambda_{\IMBH} = \lambda - t\sqrt{GM_{\IMBH}a_{\IMBH}^{-3}}$ 
(which is just $\varphi$ in the co-planar case).\footnote{While $J_r$ is dimension-less, the entire treatment can be repeated using $m\sqrt{G\MSMBH a_\IMBH}J_r$ instead, which does have units of action.} We keep the $\hat{\mathbf{z}}$-component of the angular momentum and the argument of the ascending node (together comprising $\hat{\L}$) as the other dynamical phase-space variables -- they are not involved in this resonance, so need not concern us at the moment. As the transformation $(J,\lambda) \mapsto (J_r,\lambda_r)$ is a time-dependent canonical transformation, the Hamiltonian changes by $G\MSMBH mJ_r/a_{\IMBH}$, and becomes
\begin{align}
    H \mapsto& \frac{G\MSMBH m}{a_{\IMBH}}\left[ - \frac{1}{2(1+J_r)^2} - \frac{\eps}{\sqrt{1+\alpha^2-2\alpha \cos\varphi}}\right] \nonumber\\&- \frac{G\MSMBH m}{a_{\IMBH}}J_r - \frac{G\MSMBH m}{2a_{\IMBH}}.
\end{align}
We henceforth ignore the last term, as it is a constant. 
Expanding the new Hamiltonian to leading order in $J_r$, we thus have 
\begin{align}\label{eqn:resonant Hamiltonian}
    H &\mapsto \frac{G\MSMBH m}{a_{\IMBH}} H_r \equiv \frac{G\MSMBH m}{a_{\IMBH}}\left[\frac{3}{2}J_r^2 -  \frac{\eps}{\sqrt{1+\alpha^2-2\alpha \cos\varphi(\lambda_r,\theta,\theta_{\IMBH})}}\right] \nonumber\\&\approx \frac{G\MSMBH m}{a_{\IMBH}}\left[\frac{3}{2}J_r^2 -  \frac{\eps}{\sqrt{2}\sqrt{(1 + 2J_r)(1-\cos\varphi) + (3-\cos\varphi) J_r^2}}\right].
\end{align}
The level set of $H_r$ are plotted in Figure~\ref{fig:separatrix}.
The resonance is at $(\varphi, J_r) = (0,0)$; there, $\cos \varphi = 1$, so the potential term diverges at first order in $J_r$. This requires an expansion to second order in $J_r$ in the denominator, which shows that in fact this term scales like $\eps/J_r$ near the resonance. The width of the resonant region is obtained by studying the level set of $H_r$: libration occurs when the star is effectively bound to the IMBH -- when $\varphi$ is bounded, i.e.~for $H_r \leq -\eps/2$ -- and circulation occurs when $H_r \geq -\eps/2$; the line $H_r = -\eps/2$ is thus the separatrix. The maximum deviation of $\abs{J_r}$ from $0$ inside the resonant region is at $\varphi = 0$, given by $\left(\eps/3\right)^{1/3}$ -- this is just obtained by setting $H_r = -\eps/2$ and $\varphi = 0$ in equation \eqref{eqn:resonant Hamiltonian} and solving for $J_r$ -- so that the width of the resonance is 
\begin{equation}\label{eqn: epsilon to 1 third}
    \Delta J_{r,\max} = 2\left(\frac{m_{\IMBH}}{3\MSMBH}\right)^{1/3},
\end{equation}
corresponding to 
\begin{equation}
    \frac{\Delta a_{\max}}{a} = \left(\frac{m_{\IMBH}}{3\MSMBH}\right)^{1/3}. 
\end{equation}
This is in fact just the standard Hill radius of the IMBH. 

Just outside the separatrix, $J_r$ reaches its maximum at $\varphi = 0$, and its minimum at $\pm \pi$. These are given by 
\begin{equation}
    \frac{3}{2}J_{r,\max}^2 - \frac{3}{2}J_{r,\min}^2 =  \frac{\eps}{\abs{1-\alpha}} - \frac{\eps}{1+\alpha},
\end{equation}
where the right-hand side is evaluated at some value $J_r$ between $J_{r,\min}$ and $J_{r,\max}$ -- it does not matter which, to leading order in $\eps$. Near the resonance $1+\alpha \approx 2+2J_{r}$ and $\abs{\alpha-1} \approx 2\abs{J_r}$.
After some algebra we find for the circulating case i.e.~not bound to the IMBH, with $\Delta a<\Delta a_{\max}$, the IMBH drives a systematic (oscillatory) variation of the semi-major axis relative to the SMBH, with amplitude
\begin{equation}\label{eqn:width circulation}
    \frac{\Delta J_r}{J_r} = 2\frac{\Delta a}{a} = \frac{8\eps\min\{1,\alpha\}}{3\abs{1-\alpha^2}(1-\alpha)^2}.
\end{equation} 
Observe, that if we set $\abs{1-\alpha^2} = (\eps/3)^{1/3}$ in equation \eqref{eqn:width circulation}, we find $\Delta J_r = (\eps/3)^{1/3}$ -- exactly half of the librating variation. The factor of two is expected, $J_r$ does not cross $0$ outside the separatrix. 

\begin{figure*} 
    \centering
    \includegraphics[width=0.48\linewidth]{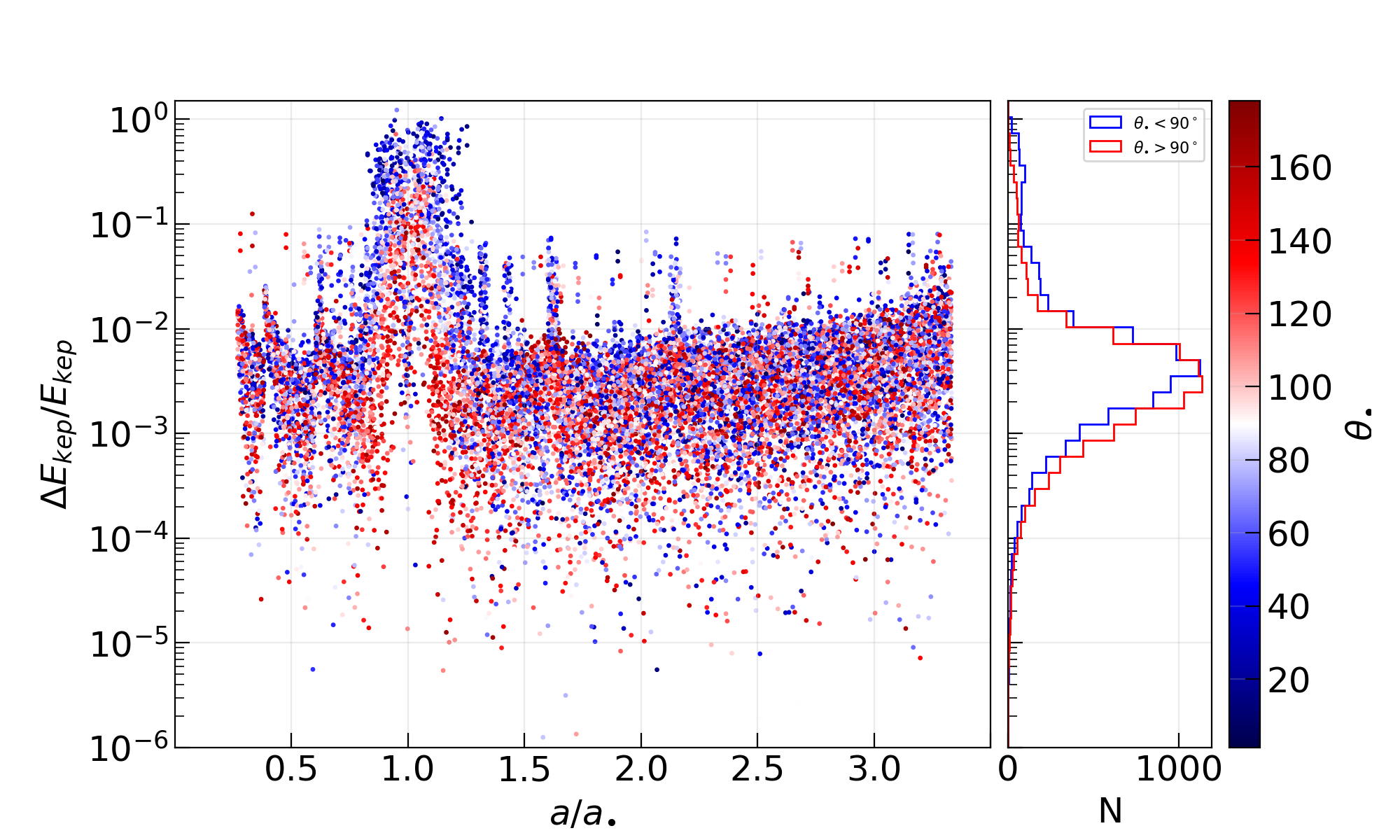}
    \includegraphics[width=0.48\linewidth]{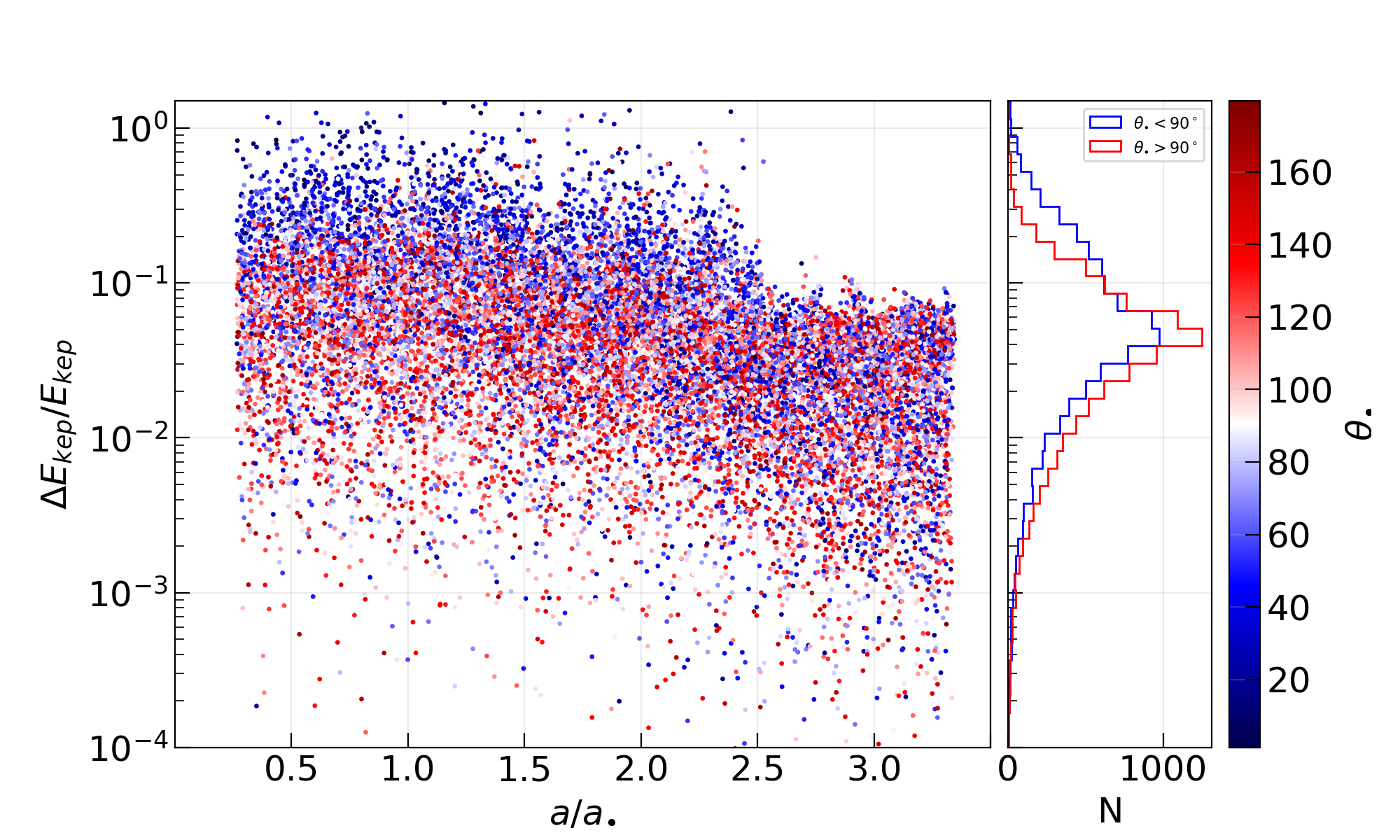}
    \caption{Relative changes of Keplerian energies for all pairs from $N$-body simulations over 10 Myr. Physical units are the same as in Fig.~\ref{fig:torques}. The systematic increase in energy with distance ratio is due to the assumed Plummer potential which represents the contribution of stars. \emph{Left}: circular orbits; \emph{right}: eccentric orbits, with $e_\IMBH = 0.7$ and $e = 0.3$ for the stars.}
    \label{fig:de}
\end{figure*}

The left panel of Figure~\ref{fig:de} exemplifies this, by showing the energy changes after $10$ Myr of evolution, for $N$-body simulations. We ran $10^4$ four-body problems with $m_\IMBH=2000\msun$, $a_\IMBH=0.15$~pc, $m_i=m_j=10\msun$ embedded in a Plummer potential with mass $M_\mathrm{pl}=10^5\msun$ and scale radius $a_\mathrm{pl}=0.2$~pc.  
We see indeed a sharp peak, of width $\sim \eps^{1/3}$, about $a=a_\IMBH$. However, away from this resonance, the assumption that semi-major axes are conserved (which underlies the entire VRR formalism) does hold to a good accuracy. 

For eccentric orbits, there will be a multitude of resonances, which preclude analytical treatment. The condition found above for there to be a large energy change is to be inside the resonance, delineated by the Hill radius. For the eccentric case, we expect every semi-major axis such that the orbits of the stars and the IMBH overlap, to be inside some resonance, so if
$r_{\rm a,\IMBH} > r_{\rm p}$ and $r_{\rm p,\IMBH} < r_{\rm a}$, then we expect there to be a significant energy exchange, leading to a breakdown of the averaged theory. Then, each value of $a$ that falls between $a_\IMBH(1+e_\IMBH)/(1-e)$ and $a_\IMBH(1-e_\IMBH)/(1+e)$, there should be a $\Delta a \sim \eps^{1/3}$. We test that in the right panel of Figure~\ref{fig:de}, where $e_\IMBH = 0.7$ and $e = 0.3$. The condition above implies that every $a \in [3/13,17/7]a_\IMBH \approx [0.23,2.43]a_\IMBH$ is strongly affected, which agrees with the simulation data. We also see in this figure, that while for the circular case there exists only a slight imbalance in energy changes as a function of inclination, this is much more pronounced in the eccentric case. Qualitatively, this arises from the known tendency of retrograde orbits to be more stable than prograde ones \citep{MardlingAarseth2001}.

\subsection{Mass precession}

\begin{figure*}
    \centering
    \includegraphics[width=\linewidth]{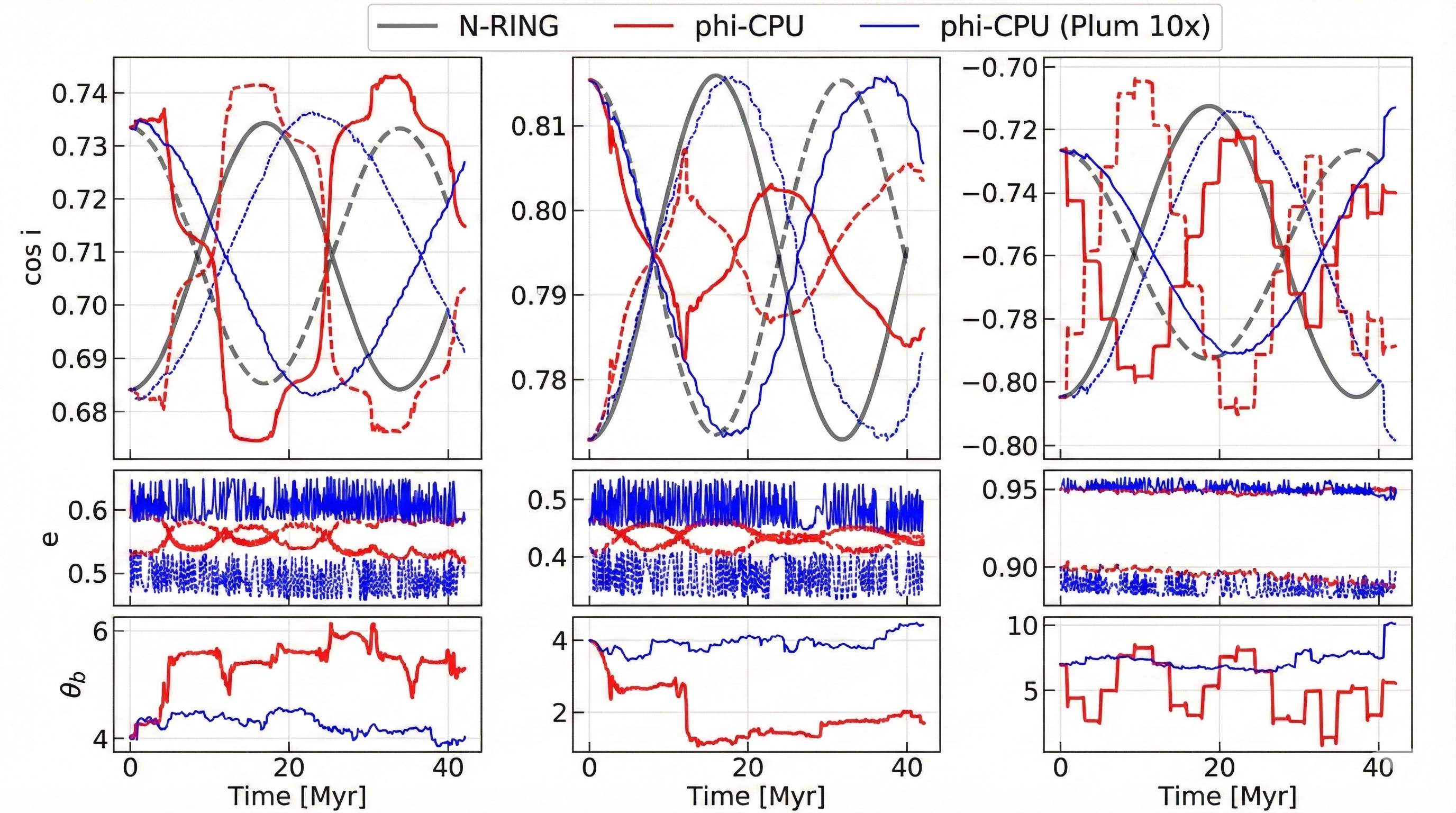}
    \caption{\emph{Top} panels: The time evolution of inclination angles of a single pair of stars with respect to an arbitrary plane (arranged similarly to Fig.~\ref{fig:noIMBH}, without an IMBH) for two-ring (integrated with \nring{}) versus three-body (integrated with \phicpu) systems. \emph{Middle} panels: Evolution of eccentricities for each pair. \emph{Bottom} panels: Evolution of relative inclination between components of each pair. Solid lines in the two upper rows represent component 1 of the pair; dashed lines represent component 2. Grey curves correspond to the \nring{} code; red curves correspond to \phicpu with the Plummer potential mass $M_\mathrm{pl}=10^5M_{\odot}$; blue curves correspond to $M_\mathrm{pl}=10^6M_{\odot}$. Different columns show a simulations with different initial conditions. The figure highlights differences between the codes, showing agreement between \nring{} and the high-Plummer mass models, but differences with respect to the lower Plummer mass models.}
    \label{fig:two-ring}
\end{figure*}

VRR presupposes that the time-scale for the evolution of the directions of angular-momentum vectors is much shorter than the time-scale on which their magnitudes evolve. This is equivalent to assuming that the arguments of pericentre evolve much more rapidly than $\hat{\L}$, and hence that one may average over them, when deriving the equations of motion of $\hat{\L}$. However, for this to be valid, the mass interior to the orbits (besides the SMBH) must be large enough to drive such a rapid pericentre precession. Let us gauge the validity of such an assumption, for parameters relevant to the Milky Way's nucleus. 

In the \phicpu code, we mimic mass precession by introducing a Plummer potential with a scale radius of $a_\mathrm{pl} = 0.2$~pc and a mass $M_\mathrm{pl}=10^5\msun$; additionally, we perform simulations with $M_\mathrm{pl}=10^6\msun$ to examine differences arising from an increased enclosed mass. To pinpoint these differences, we compare the pairwise evolution of two rings (for which an analytical solution is known) integrated with \nring, to the evolution of equivalent three-body systems (the central SMBH and two stars) integrated with \phicpu. For the \phicpu~simulations, we use two values for the enclosed Plummer mass as described above. Figure~\ref{fig:two-ring} shows the evolution of inclination angles of each component of the pair (top panels), eccentricities of the components (middle panels), and relative inclination between the components (bottom panels). Eccentricities and relative inclinations are shown only for \phicpu, as in \nring~these quantities are conserved by construction. Each column corresponds to one simulation, featuring two rings in the case of \nring~and a three-body configuration for \phicpu. The \nring~simulations match the analytical solution precisely, exhibiting smooth precession of angular momentum vectors (grey curves in the figure), whereas in the \phicpu~simulations, especially with lower enclosed mass, noticeable deviations occur. These deviations are also accompanied by eccentricity variations.

A comparison between the red ($M_\mathrm{pl}=10^5\msun$) and blue ($M_\mathrm{pl}=10^6\msun$) curves indicates that a higher Plummer mass reduces deviations observed in the red curves, effectively averaging them out. Models with higher Plummer mass also exhibit nearly constant relative inclinations between the pair components. Interestingly, the high-eccentricity model (right column), in case of lower Plummer mass, displays a step-like evolution, characterised by abrupt changes at constant time-intervals corresponding to the apsidal-precession time, but these step-like oscillations are averaged-out in the model with higher mass. In summary, a higher enclosed mass tends to average out deviations, producing results closer to those obtained with \nring, though some differences in precession periods remain.

The chosen Plummer parameters ($M_\mathrm{pl}=10^5$--$10^6\msun$ and $a_\mathrm{pl}=0.2\,\mathrm{pc}$) are representative of the stellar mass distribution within the central parsec of the Milky Way. Observational studies suggest that the stellar mass enclosed within the central $0.5,\mathrm{pc}$ is of the order of a few $10^5\msun$, while within $1\,\mathrm{pc}$ it reaches approximately $10^6\msun$ \citep[e.g.][]{Schodel2018,Gallego-Cano2018}. Thus, our adopted values are realistic and provide a suitable framework to assess the validity of the VRR approximation in conditions relevant to the Galactic centre. We see that the innermost regions of the Galactic centre might exhibit deviations induced by relatively lower enclosed mass. Additionally, it is important to note that stellar densities in the Milky Way nucleus typically follow a power-law cusp rather than a Plummer profile; nonetheless, the chosen Plummer parameters closely approximate the Bahcall–Wolf density distribution over most of the considered radial range.

In conclusion, our numerical experiments indicate that three-body simulations (\phicpu) exhibit deviations from the idealised VRR predictions due to finite enclosed-mass effects. While these deviations are noticeable, the overall evolutionary trend remains dominated by the dynamics described by the VRR formalism. Furthermore, increasing the enclosed mass significantly reduces these deviations, reinforcing the applicability and validity of the VRR approximation under realistic mass conditions relevant to the central parsec of the Milky Way.

\section{Applications to the Galactic Centre}
\label{sec:applications_to_GC}

\begin{figure*} 
    \centering
    \includegraphics[width=0.495\linewidth]{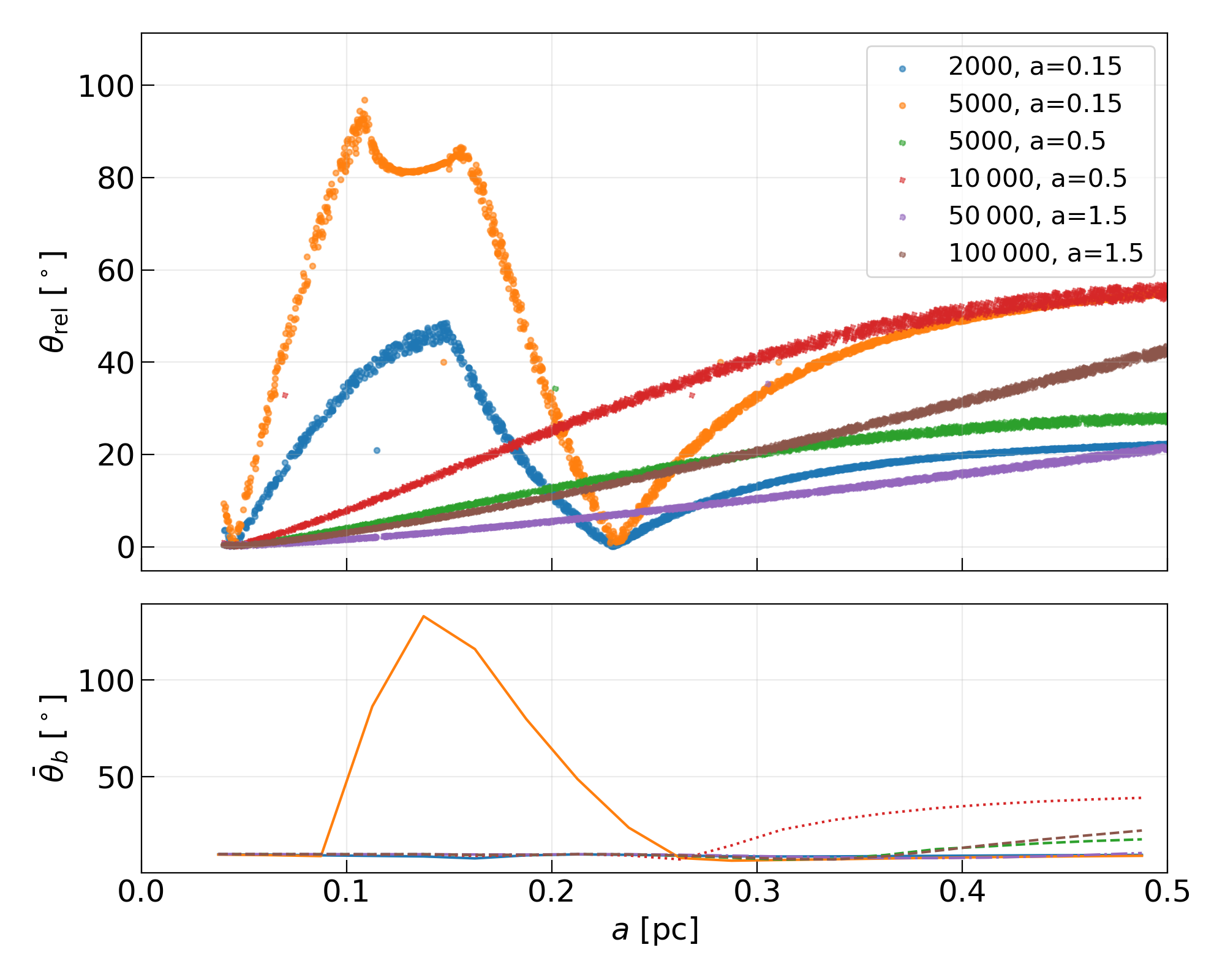}
    \includegraphics[width=0.495\linewidth]{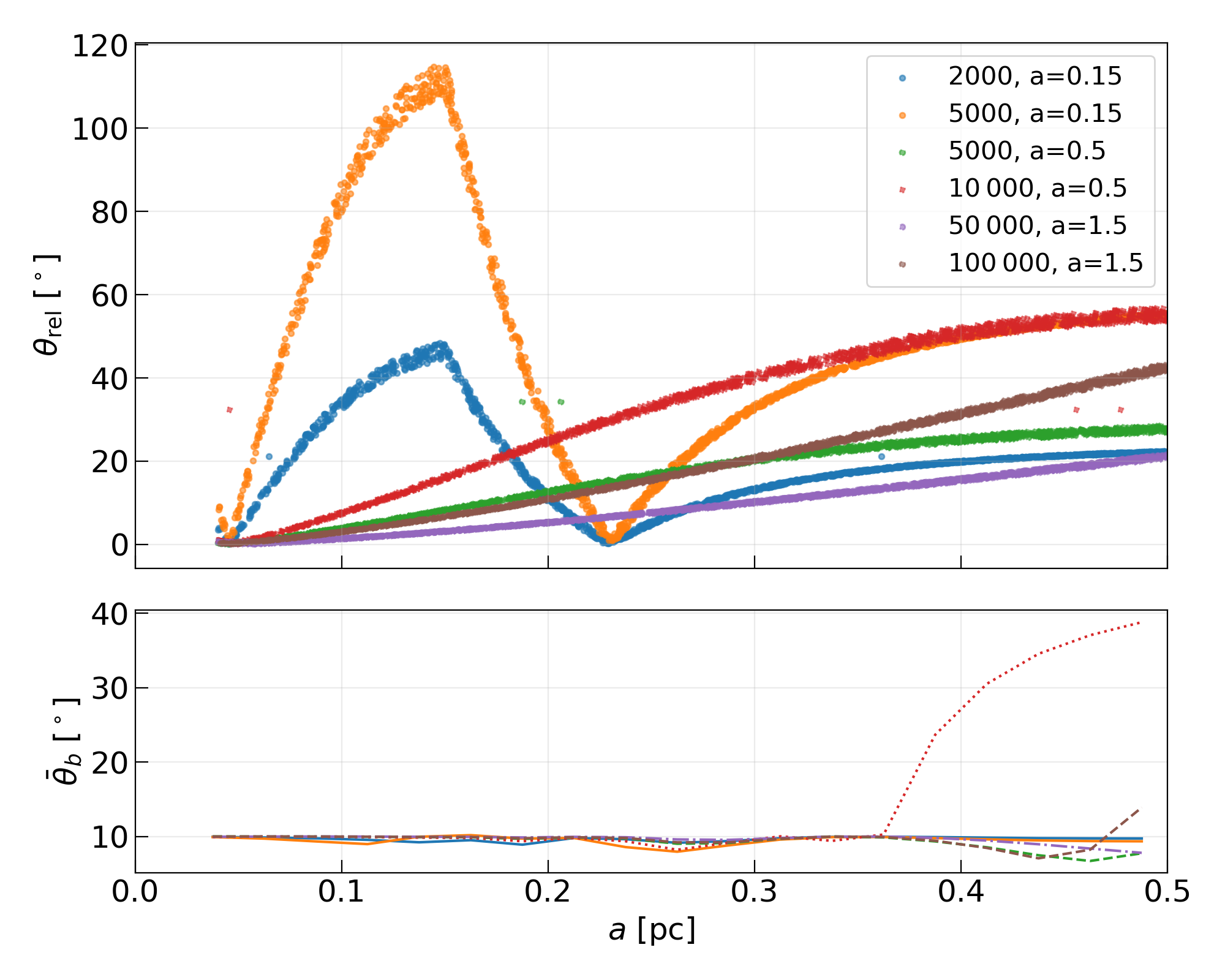}
    \caption{Relative inclination angle of the pair with respect to the inner ($a < 0.05$~pc) angular momentum vector of all pairs as a function of the semi-major axis of the pair (upper panels). Average inclination of the pair (lower panels). Both panels show results of \nring simulations for the time of 10 Myr of initially coupled pairs with relative inclination of $10^\circ$ and initial inclination angle with the perturber $\theta_\IMBH \simeq 65^\circ$. Solid lines in the lower panels correspond to an IMBH with semi-major axis $a_\IMBH=0.15$ pc; dashed and dotted lines correspond to a perturber with $a_\IMBH=0.5$ pc and $a_\IMBH=1.5$ pc, respectively, and masses according to the legend. In all cases, eccentricities of pairs and the perturber were set to $e_s = e_\IMBH = 0.3$. \emph{Left:} $m_i+m_j=100\msun$. \emph{Right:} $m_i+m_j=200\msun$. See text for details.}
    \label{fig:gc_app}
\end{figure*}

Let us explore a simple consequence of the stability criterion derived above for the young stellar disc in the Milky Way. In order to derive order-of-magnitude estimates for IMBH-induced nodal precession in a young stellar disc, we treat the disc as an ensemble of independent ``fragments''. Each fragment is comprised of two concentric rings that mutually torque one another and jointly experience the torque from an IMBH. Given that the internal interaction within a fragment is expected to be much larger due to their proximity than the interaction between different fragments, we simply assume that the different fragments do not interact with each other. Hence, we analyse each two-ring system and the perturber separately. We then ask: if an IMBH lies either inside or outside the disc, what conditions are necessary on its mass and orbital parameters for it to disrupt the disc and drive an initially thin, uni-directional disc to retrograde orbits? In particular, we identify (i) the critical distance from the IMBH at which a fragment's ring-pair becomes dynamically ``disrupted'', (ii) the maximum relative inclination of each pair (with respect to the inner angular-momentum vector of all pairs) that the IMBH can induce, and (iii) the relative inclination of each pair as a function of distance.

For all pairs, we set eccentricities $e_s = 0.3\pm 0.03$, and we explore a range of semi-major axes $0.04 \leq a \leq 0.5~\text{pc}$. This parameter set mimics that of the young stars in the central parsec of the Galaxy \citep{vonFellenberg2022}. For the perturber, we set the eccentricity to $e_\IMBH = 0.3$ but vary its semi-major axis. The discrete values chosen are: $a_\IMBH=0.15$ pc, the projected distance of the IRS13 association, potentially hosting an IMBH \citep{Tsuboi2017}; $a_\IMBH=0.5$ pc, the possible 3D distance of IRS13 \citep{Tsuboi2020}; and $a_\IMBH=1.5$ pc, an estimated distance to the circumnuclear disc (e.g.~\citealt{Oka2011}). Additionally, we vary the mass of the perturber from $2\times10^3$ to $10^5~\msun$, placing lighter perturbers closer to the SMBH to operate within the allowed range of masses for the putative IMBHs \citep{Lommen_Backer2001,GualandrisMerritt2009,Gualandris+2010,Naoz2020,Gravity2020, Gravity2023}. 

An important parameter is the mass of the fragment (i.e. a single clump) relative to the IMBH mass; we thus choose two different clump masses: $m_i+m_j=100\msun$ and $m_i+m_j=200\msun$, with $m_i=m_j$ in both cases. The clump mass is set by the number of particles with the strongest VRR coupling coefficients; in general, it may be defined by the number of objects with overlapping and marginally overlapping orbits and may depend on the position of objects within the disc. However, in our toy model we adopt the two values above for simplicity. Finally, we define the innermost pairs as the ones with semi-major axes $0.04 \leq a \leq 0.05$~pc. Having set all the parameters we can use the disruption criterion in Eq.~\eqref{eq:synchronous} which tells us that the pairs with $m_i+m_j=100\msun$ will be decoupled only for the IMBH of $5\times10^3\msun$ and for some outer perturbers with $a_\IMBH=0.5$ and 1.5~pc. On the other hand, while clumps with $m_i+m_j=200\msun$ are expected to remain coupled for all inner perturbers with the parameters listed above, they can be decoupled by the IMBH of $10^4\msun$ with $a_\IMBH=0.5$~pc. Additionally, if we were to include interactions between different fragments, any clump whose orbit overlaps or marginally overlaps with that of the IMBH would be decoupled from the innermost clumps according to the general condition in Eq.~\eqref{eqn:fundamental criterion} (which must be used for large orbital separations between clumps). Because we ignore such fragment–fragment interactions in our toy model, we instead treat the innermost region as precessing independently about the IMBH and measure the inclination of all other clumps with respect to this innermost disc. Thus, it transpires that for this toy model, the stellar disc is expected to be stable in most of the parameter space -- the clumps should be coupled to each other but at the same time they should be de-coupled with respect to the inner clumps. Therefore, by running \nring simulations we can measure relative inclinations of the clumps with respect to the inner disc and relative inclination between the clumps at 10~Myr.

We ran a set of isolated three-ring problems in \nring with initial conditions adjusted to mimic the Galactic centre. We initialised all pairs with relative inclinations of $\theta_b = 10^\circ$ and inclination with respect to the perturber $\theta_\IMBH\simeq65^\circ$. We chose this inclination because, in the case of disruption, the maximum relative angle $\theta_1 + \theta_2=130^\circ$ exceeds $100^\circ$, pushing the corresponding pairs into retrograde orbits relative to the inner objects. 
For each set of initial conditions, we perform $10^4$ three-ring simulations.

To interpret the results, we stacked all pairs in a single snapshot at $t = 10~\text{Myr}$ and measured their relative inclinations $\theta_b$, as well as the relative inclination $\theta_\mathrm{rel}$ of each pair's total angular momentum $\L_i+\L_j$ with respect to the net angular momentum of the innermost pairs (semi-major axes $0.04 \leq a \leq 0.05$~pc). 

Figure~\ref{fig:gc_app} shows the resulting relative angle $\theta_\mathrm{rel}$ as a function of semi-major axis (upper panels) and the pairs' average inclination, defined as $\bar{\theta}_b = \arccos(\langle \cos\theta_b\rangle)$ (lower panels), indicating regions where pairs are disrupted or remain coupled. The left panels correspond to fragments with total mass $100\msun$, and the right panels correspond to fragments with mass $200\msun$. To obtain statistics, we bin the data with bin width 0.025~pc, resulting in approximately 115 objects per bin in the inner region and up to about 635 objects per bin in the outer region.

The simulation results indicate that, for clump masses of $100\msun$, an IMBH of mass $5\times10^3\msun$ at $a_\IMBH=0.15~\text{pc}$ pushes nearby pairs into retrograde orbits relative to the inner disc (orange curve, upper left panel) and significantly disrupts this region (orange curve, lower left panel). However, a stellar disc with clump masses of $200\msun$ maintains its thickness (lower right panel, nearly constant lines), while simultaneously re-orienting regions around $0.15~\text{pc}$ into retrograde orbits. A lighter IMBH at this distance can reorient the disc less dramatically, reaching about $50^\circ$ relative inclination within 10 Myr. These cases resemble the observed distributions of young stars, notably in \citet{Bartko2009} -- the low mass clumps produce many particles outside of the disc plane while the more massive clumps tend to preserve the original thickness of the disc. This suggests that in more realistic scenarios characterized by a variable stellar mass function and diverse orbital configurations, it is natural to expect a superposition of these effects resulting in complex disc structures, as recently demonstrated in observational studies \citep{vonFellenberg2022, Siyao2023}.

More distant and massive IMBHs warp the stellar disc, but only an IMBH of $10^4\msun$ significantly disrupts it, increasing the thickness to $\sim40^\circ$ in 10 Myr.

In summary, our toy model favours an IMBH mass of about $5\times10^3\msun$. However, this model treats the disc fragment as two rings, neglecting its interaction with the rest of the disc, and also does not include torques due to fluctuating anisotropy in the nuclear star cluster, which could be significant on the relevant time-scales (e.g. \citealt{Panamarev2022}), nor does it account for the inward migration of the IMBH induced by Chandrasekhar dynamical friction \citep{Chandra1943}, or an orbital alignment due to resonant dynamical friction \citep{Szolgyen2021,Ginat2022}. 
Let us estimate the effect that dynamical friction with the nuclear cluster would have on the IMBH, and see if it could cause it to lose energy and sink into the SMBH. The (non-resonant) Chandrasekhar dynamical friction force on an IMBH of mass $m_\IMBH$ moving with velocity $\mathbf{v}$ through a background of stars of density $\rho$ and one-dimensional velocity dispersion $\sigma$ can be written as
\begin{equation}
    \mathbf{F}_{\rm df} = -\,4\pi\,G^2\,m^2_\IMBH \,\ln\Lambda\,\frac{\rho}{v^3}\,f(X)\,\mathbf{v}\,,
\end{equation}
where $v = |\mathbf{v}|$, $\ln\Lambda$ is the Coulomb logarithm, $X \equiv v/(\sqrt{2}\,\sigma)$, and $f(X)$ is a dimensionless function of order unity that encodes the dependence on the ratio of the IMBH speed to the stellar velocity dispersion \citep{Binney}. 
For a circular Keplerian orbit of radius $a_0$ around the SMBH, with $v \simeq \sqrt{G M/a_0}$ and $\sigma \simeq v/\sqrt{2}$, this force implies a local dynamical-friction time-scale $t_{\rm df,loc}(a_0) \equiv a_0/|\dot a|_{a_0}$ which can be compared to the local two-body relaxation time $t_r(a_0) \simeq 0.34\,\sigma^3/(G^2 m_\star \rho \ln\Lambda)$, where $m_\star$ is the typical stellar mass. Using these expressions ($f(X)\approx0.428$) one finds
\begin{equation}
    t_{\rm df,loc}(a_0)\simeq 0.77\frac{m_\star}{m_\IMBH}t_r(a_0),
\end{equation}
so that for $m_\IMBH \gg m_\star$ the IMBH inspirals on a time-scale much shorter than the local relaxation time. Taking $t_r \approx 10$ Gyr (see e.g. \citealt{Kocsis2011, Panamarev2022}) at the edge of the disc ($a_0 \simeq 0.5$ pc), we find that for IMBHs with masses $m_\IMBH \ge 10^4\msun$ the dynamical friction time drops below $1$ Myr, which is significantly shorter than the age of the disc ($6$--$10$ Myr; \citealt{Bartko2010}). We therefore conclude that such massive black holes are unlikely to exist within the disc.

Including the additional effects described above may reveal that lower-mass IMBHs can also effectively reorient the stellar disc into retrograde orbits within $\sim 10$ Myr. Although our simplified model provides useful order-of-magnitude estimates, it can neither rule out the existence of IMBHs, nor prove them -- merely point to the parameter values that merit a more careful study. The most reliable way to explore these complex dynamics comprehensively is through direct $N$-body simulations. Thus, the favoured parameters identified here serve as robust initial conditions for future, large-scale numerical investigations aimed at uncovering the precise role of IMBHs in shaping the distribution of stars in the Galactic centre.

Additionally, gas forces may affect the dynamics at early times. The disc may become warped in the gas phase, before stars have formed \citep{LevinBeloborodov2003,Nealon2015}, interaction of stars with the gaseous disc may lead to orbital alignment, circularization and semimajor axis decay \citep{KenEtAl2016, Panamarev2018} and/or the gas could affect the dynamics of the IMBH \citep{Fabj2020}. However, coupling gas physics with stellar dynamics requires computationally expensive hydrodynamical simulations integrated with $N$-body gravity, a task of significant complexity that demands a dedicated investigation and is therefore beyond the scope of this work.

\section{Summary and Discussion}
\label{sec:summary}

We have pointed out a novel phenomenon in multi-body gravitating systems subject to a predominantly spherical potential in which the bodies execute planar orbits. In these systems the orbital planes, represented by the angular momentum vectors, reorient rapidly due to vector resonant relaxation (VRR). We have shown that in these systems angular momentum vectors may exhibit bound long-lived sub-clusters in angular momentum space, which evolve together due to their strong VRR coupling. In particular, we examined two bodies on nearly coplanar orbits with initially nearly parallel angular momentum vectors, that initially precess about their net angular momentum, and experience perturbations from a distant fourth body whose angular momentum dominates the system. We derived the conditions under which the initially coupled pair becomes disrupted by the perturber, causing their angular-momentum vectors to precess independently around the perturber's angular-momentum vector. We investigated various orbital configurations, including circular orbits (Sec.~\ref{subsec:circular}), eccentric radially non-overlapping orbits, and eccentric radially overlapping orbits (Sec.~\ref{subsec:elliptical orbits}). A general stability criterion \eqref{eqn:fundamental criterion} was established and summarised by Eq.~\eqref{eq:synchronous}, validated through numerical simulations, as shown in Fig.~\ref{fig:rel_torques}. Additionally, we derived analytical expressions in the quadrupole approximation and for cases involving radially overlapping orbits. These analytical results were further numerically tested using the \nring{} code, simulating the double-orbit-averaged equations of motion in the VRR formalism, and using the direct $N$-body code \phicpu (Fig.~\ref{fig:torques_all}). Comparing the results obtained with these different methodologies, we quantified the validity of the VRR formalism when applied to more realistic settings (Figs~\ref{fig:de} and \ref{fig:two-ring}).

We applied this stability criterion to construct a simplified model representing the disc of young massive stars in the Galactic centre interacting with a massive external perturber, such as a putative IMBH or the circumnuclear gaseous torus (the circumnuclear disc). Our model provides order-of-magnitude estimates for a possible perturber’s orbital parameters required to produce specific observational features in the angular-momentum distribution of disc stars, notably the clustering of angular momentum vectors and forming a retrograde structure. These results also identify the possible parameters that can serve as initial conditions for large-scale direct $N$-body simulations of the Galactic centre, analogous to those performed by \citet{Panamarev2022}.

An important assumption in our analysis is that the IMBH dominates the system's angular momentum. However, if the stellar disc mass is comparable to or greater than that of the IMBH, their angular momenta become similar in magnitude, altering the trajectories of angular-momentum vectors on the unit sphere (Fig.~\ref{fig:sphere_demo}). For instance, if the net angular momentum vector of the stellar disc is comparable to that of the IMBH, placing it at the pole in Fig.~\ref{fig:sphere_demo} would allow for large relative inclinations within disrupted angular momentum pairs, especially when the IMBH is initially placed on a retrograde orbit relative to the disc. Furthermore, initially prograde IMBH orbits are expected to produce lower relative inclinations.

Our simplified models hint that an IMBH might exist with a mass in the range of $2$--$5\times10^3\msun$, orbiting at a semi-major axis of approximately $0.15~\text{pc}$ and moderate eccentricity (we adopted $e_\IMBH=0.3$). Such an IMBH configuration may produce the observed retrograde structures from an initially flat stellar disc formed in a single star formation event from a gaseous accretion disc. 
Furthermore, the model favours a relatively high initial inclination of the IMBH with respect to the stellar disc -- exceeding $50^\circ$ -- in order to efficiently produce the retrograde disc structure observed in the Galactic centre within 10 Myr consistent with the estimated age of the young stars in the Galactic centre \citep{Bartko2010, Genzel2010, Habibi2017}. 
In the Milky Way, the net angular momentum of the stellar disc may exceed that of the IMBH. However, even in such cases, the IMBH may still dominate locally within the region of overlapping orbits. This local dominance could validate our model assumptions in that region, thereby allowing the formation of the retrograde feature.

A comprehensive exploration of disc stability under the influence of multiple perturbers, including IMBHs and the surrounding nuclear star cluster, is beyond the scope of the current work. Such a detailed investigation, including direct $N$-body simulations that self-consistently incorporate these additional effects, is deferred to future work.

\section{Acknowledgements}
We thank the referee for useful comments and suggestions. We are grateful to Peter Berczik for providing us with the CPU version of $\phi$-GPU.  
This work was supported by the Science and Technology Facilities Council grant number ST/W000903/1, and by a Leverhulme Trust International Professorship Grant (No. LIP-2020-014). We acknowledge the support of the  Science Committee of the Ministry of Education and Science of the Republic of Kazakhstan (Grant No. AP22787256). The work of Y.B.G.~was partly supported by a Simons Investigator Award to A.A.~Schekochihin.

\section*{Data Availability}
The data underlying this article will be shared on reasonable request to the corresponding author.

\bibliography{thesis,encounters}

\appendix
\onecolumn


\section{Legendre sums}
\label{App:Legendre}

In this appendix we provide properties of Legendre polynomials from the mathematical literature \citep[e.g.][]{DLMF}, which we use in this paper, for the readers' convenience.  

The VRR Hamiltonian in Eq.~\eqref{eq:H_VRR} leads to equations of motion (Eq.~\eqref{eq:EOM}) containing infinite sums of the form \(\sum_\ell [P_{\ell}(0)]^2P_\ell(x)\,s_{ij\ell}\alpha^\ell\), where $[P_{\ell}(0)]^2$ is given by Eq.~\eqref{eq:pnzero}. For circular orbits this is  \(\sum_\ell [P_{\ell}(0)]^2P_\ell(x)\,z^\ell\), where $z=r_\mathrm{in}/r_\mathrm{out}$. \citet{Kocsis2015} showed (see their Appendix B5) that for both eccentric orbits 
\begin{equation}
s_{ij\ell}\alpha^\ell \rightarrow \frac{1}{\ell}\left(\frac{r_\mathrm{a,in}}{r_\mathrm{p,out}}\right)^\ell
\end{equation}
for large $\ell$. Using Stirling's approximation to the Gamma function, one can show that for large $\ell$
\begin{equation}
\label{eq:P_2n0}
\left[P_{2\ell}(0)\right]^2 \sim 
\frac{1}{\pi \ell}\left( 1 - \frac{1}{4\ell} + \frac{5}{32 \ell^{2}} + O\!\left(\ell^{-3}\right) \right).
\end{equation}

Thus, we will be dealing with the sums of the form
\(\sum_\ell \ell^{-n}P_\ell(x)\,z^\ell\).  For \(|z|\le1\) (i.e.\ for both non-overlapping and overlapping orbits) these sums admit closed-form expressions via the standard Legendre generating function. We briefly outline the method here:

\begin{enumerate}
  \item Start from the generating function
 \begin{equation}
     G(x,z)=\sum_{\ell=0}^\infty P_\ell(x)\,z^\ell
    =(1-2 x z+z^2)^{-1/2}.
 \end{equation}
  \item Subtract the constant term, divide by $z$ and integrate with respect to $z$:
\begin{equation}
    \sum_{\ell=1}^\infty \frac{P_\ell(x)}{\ell}\,z^\ell
      =\int_{0}^{z}\frac{G(x,\xi)-1}{\xi}\,\mathrm{d}\xi.    
\end{equation}
  \item Divide by $z$ and integrate a second time:
 \begin{equation}
    f(x;z)=\sum_{\ell=1}^\infty \frac{P_\ell(x)}{\ell^2}\,z^\ell
      =\int_{0}^{z}\frac{1}{\eta}
       \int_{0}^{\eta}\frac{G(x,\xi)-1}{\xi}\,\mathrm{d}\xi\,\mathrm{d}\eta.
\end{equation}
  \item Differentiate with respect to $x$.  Moving the $x$-derivative inside the integrals introduces $P'_\ell(x)$ under the sum.
  \item To obtain expressions for even or odd $\ell$ respectively, take the even/odd part  of the function with respect to $z$, i.e. $\frac12[f'(z)\pm f'(-z)]$ yields linear combinations of $z^\ell$ with only even or only odd $\ell$.
  \item Similarly, if we integrate the generating function first and then divide by $z$ we get expressions of the form
 \begin{align}
    &\sum_{\ell=1}^\infty \frac{P_\ell(x)}{\ell + 1}\,z^\ell
      =\frac{1}{z}\int_{0}^{z} G(x,\xi)\,\mathrm{d}\xi.\\
    &\sum_{\ell=1}^\infty \frac{P_\ell(x)}{(\ell+1)(\ell+2)}\,z^\ell
      =\frac{1}{z^2}\int_{0}^{z}
       \int_{0}^{\eta} G(x,\xi)\,\mathrm{d}\xi\,\mathrm{d}\eta. 
\end{align}
The integrals on the right hand side of all of these expressions simplify analytically.
  
\end{enumerate}

Introducing
\[
  x=\cos\theta=\mathbf{L}_i\cdot\mathbf{L}_j,\quad
  X=\sqrt{1-2 x z+z^2},\quad
  Y=\sqrt{1+2 x z+z^2},
\]
we get the general results
\begin{align}
  &\sum_{\ell=1}^{\infty}\frac{P'_{\ell}(x)}{\ell^2}\,z^{\ell}
    = \frac{1}{1-x}\ln\left(\frac{1+X+z}{2}\right)
     -\frac{1}{1+x}\ln\left(\frac{1+X-z}{2}\right)\,,\\
  &\sum_{\ell=2,4}^{\infty}\frac{P'_{\ell}(x)}{\ell^2}\,z^{\ell}
    = \frac{1/2}{1-x}\ln\left[\frac{(1+X+z)(1+Y-z)}{4}\right]
     -\frac{1/2}{1+x}\ln\left(\frac{(1+X-z)(1+Y+z)}{4}\right)\,,\\
  &\sum_{\ell=0}^{\infty}\frac{P'_{\ell}(x)}{(\ell+1)(\ell+2)}\,z^{\ell}
    = \frac{1-X}{z^2}
     +\frac{z-x}{z^2}\ln\left(\frac{z-x+X}{1-x}\right)\,,\\
  &\sum_{\ell=0,2}^{\infty}\frac{P'_{\ell}(x)}{(\ell+1)(\ell+2)}\,z^{\ell}
    = \frac{2-X-Y+2x\ln(1-x)}{2z^2}
     +\frac{z-x}{2z^2}\ln\left(z-x+X\right)
     +\frac{-z-x}{2z^2}\ln\left(-z-x+Y\right)\,.
\end{align}
For $z=1$ (overlapping orbits) this simplifies further as
\begin{align}
    &\sum_{\ell=1}^{\infty}\frac{P_{\ell}(x)}{\ell^2}
        =  \frac{\pi^2}{6}-2\chi(s)-2\ln s \ln c\,,\\
\end{align}
where $\chi(z)=\sum_{n=0}^{\infty} z^{2n+1}/(2n+1)^2$ is the Legendre-chi function which may be expressed with the dilogarithm $\mathrm{Li}_2(z)=\sum_{n=1}^{\infty}z^{n}/n^2$ as $\chi(z)=\frac12[\mathrm{Li}_2(z)-\mathrm{Li}_2(-z)]$.
\begin{align}    
    &\sum_{\ell=2,4}^{\infty}\frac{P_{\ell}(x)}{\ell^2}
        =  \frac{\pi^2}{6}-\chi(s)-\chi(c)-2\ln s \ln c\,, \\        
    &\sum_{\ell=0}^{\infty}\frac{P_{\ell}(x)}{(\ell+1)(\ell+2)}
        =  1 - 2s + 2s^2\ln\left(\frac{1+s}{s}\right)\,, \\
    &\sum_{\ell=0,2}^{\infty}\frac{P_{\ell}(x)}{(\ell+1)(\ell+2)}
        =  1 - s - c + s^2\ln\left(\frac{1+s}{s}\right) + c^2\ln\left(\frac{1+c}{c}\right)\,, \\
    &\sum_{\ell=1}^{\infty}\frac{P'_{\ell}(x)}{\ell^2}
        = \frac{\ln(1+s)}{1-x}
        -\frac{\ln(s)}{1+x} = \frac{\ln(1+s)}{2s^2}
        -\frac{\ln(s)}{2c^2}\,,\\
    &\sum_{\ell=2,4}^{\infty}\frac{P'_{\ell}(x)}{\ell^2}
        = \frac{\ln[(1+s)c]}{4s^2}
        -\frac{\ln[(1+c)s]}{4c^2}\,,\\      
  &\sum_{\ell=0}^{\infty}[P_{\ell}(0)]^2P_{\ell}(x)
        =\frac{2}{\pi}K(x) ~~\text{where $K(k)$ is the complete elliptic integral of the first kind, }
\end{align}
where $s=\sqrt{(1-x)/2}=\sin(\theta/2)$ and $c=\sqrt{(1+x)/2}=\cos(\theta/2)$. 

Taking the difference of the two types of expressions gives a formula for $\ell^{-3} P_{\ell}$ type sums. For instance
\begin{align}\label{eq:sumell3}
\sum_{\ell=1}^{\infty}\frac{P'_{\ell}(x)}{\ell^3}
        &=  \frac{-1}{6}\left[\frac{1}{s} + \frac{1}{1+s} + \left(2+\frac{1}{c^2}\right)\ln s - \left(2+\frac{1}{s^2}\right)\ln(1+s)\right] + \frac13 \sum_{\ell=1}^{\infty}\left[\frac{1}{(\ell+1)(\ell+2)}-\frac{1}{\ell^2}\left(1-\frac{3}{\ell}\right)\right]P'_{\ell}(x),
\end{align}
Here the sum on the right hand side converges very quickly, implying that it is sufficient to take the first few terms only and neglect all others, which gives a low order polynomial there.

\subsection{Integral representations for fractional–power Legendre sums}
\label{App:Legendre_fractional}

The technique outlined above may be extended to evaluate sums containing fractional powers of~$\ell$ in the denominator, such as
\[
    \sum_{\ell}\frac{P_{\ell}(x)}{\ell^{3/2}}\,z^{\ell},
    \qquad
    \sum_{\ell}\frac{P'_{\ell}(x)}{\ell^{3/2}}\,z^{\ell},
\]
which arise in the asymptotic treatment of the VRR Hamiltonian when using the Stirling approximation for
$[P_{2\ell}(0)]^2 \!\propto \!\ell^{-1}$, yielding an overall $\ell^{-3/2}$ dependence in the coupling coefficients for the case of circular-eccentric orbital configurations (cf. Appendix B5 in \citealt{Kocsis2015}).  
These series converge for all $|z|\le 1$ and $|x|<1$ but do not possess closed–form elementary expressions; however, they admit rapidly convergent integral representations obtained from the Legendre generating function.

Starting from
\begin{equation}
    G(x,q)
    =\sum_{\ell=0}^{\infty}P_{\ell}(x)\,q^{\ell}
    =(1-2xq+q^{2})^{-1/2},
    \qquad |q|<1,
\end{equation}
we recall the general identity (valid for $\Re\,\nu>0$)
\begin{equation}
    \frac{1}{\ell^{\nu}}
    =\frac{1}{\Gamma(\nu)}
     \int_{0}^{1}(-\ln t)^{\nu-1}\,t^{\,\ell-1}\mathrm{d}t,
\end{equation}
which provides an integral kernel for replacing powers of~$\ell$ by logarithmic weights.  
Applying this to the generating function gives the Abel transform representation
\begin{equation}
\label{eq:abel_gen}
    \sum_{\ell=0}^{\infty}\frac{P_{\ell}(x)}{\ell^{\nu}}\,z^{\ell}
    =\frac{1}{\Gamma(\nu)}
     \int_{0}^{1}
      \frac{(-\ln t)^{\nu-1}}{t}\,G(x,zt)\,\mathrm{d}t,
      \qquad 0<z\le1.
\end{equation}
For $\nu=\tfrac{3}{2}$ this yields the expression used in our numerical evaluation:
\begin{equation}
\label{eq:S1_abel}
    S_{1}(x,z)\equiv
    \sum_{\ell=0}^{\infty}\frac{P_{\ell}(x)}{\ell^{3/2}}\,z^{\ell}
    =\frac{2}{\sqrt{\pi}}
     \int_{0}^{1}
      \frac{\sqrt{-\ln t}}{t}\,G(x,zt)\,\mathrm{d}t.
\end{equation}
Since $P_{\ell}(0)=0$ for all odd $\ell$, the physically relevant sums involve only even degrees.  
The even–$\ell$ projection of the generating function,
\begin{equation}
\label{eq:even_proj}
    G_{\mathrm{even}}(x,q)
    =\frac{1}{2}\left[G(x,q)+G(x,-q)\right]
    =\sum_{\ell\,\mathrm{even}}P_{\ell}(x)\,q^{\ell},
\end{equation}
therefore leads to
\begin{equation}
\label{eq:S1_even}
    S_{1}^{(\mathrm{even})}(x,z)
    =\frac{2}{\sqrt{\pi}}
     \int_{0}^{1}
      \frac{\sqrt{-\ln t}}{t}
      \bigl[G_{\mathrm{even}}(x,zt)-1\bigr]
      \,\mathrm{d}t
    =\frac{2}{\sqrt{\pi}}
     \int_{0}^{1}
      \frac{\sqrt{-\ln t}}{t}
      \left[\frac{1}{2}\!\left(\frac{1}{\sqrt{1-2xzt+z^{2}t^{2}}}
      +\frac{1}{\sqrt{1+2xzt+z^{2}t^{2}}}\right)-1\right]
      \mathrm{d}t.
\end{equation}
The subtraction of unity removes the $\ell=0$ term, ensuring convergence at the lower limit.

An entirely analogous procedure applies to the series containing the derivatives $P'_{\ell}(x)$.
Differentiating the generating function with respect to $x$ gives
\begin{equation}
    \frac{\partial G(x,q)}{\partial x}
    =\sum_{\ell=0}^{\infty}P'_{\ell}(x)\,q^{\ell}
    =\frac{q}{(1-2xq+q^{2})^{3/2}}.
\end{equation}
Applying the same fractional–power kernel and even–$\ell$ projection,
\begin{equation}
\label{eq:S2_even}
    S_{2}^{(\mathrm{even})}(x,z)
    =\frac{1}{\sqrt{\pi}}\,z
     \int_{0}^{1}\!\sqrt{-\ln t}\,
     \Bigl[(1-2xzt+z^{2}t^{2})^{-3/2}
           -(1+2xzt+z^{2}t^{2})^{-3/2}\Bigr]
     \,\mathrm{d}t.
\end{equation}
Equations~\eqref{eq:S1_even} and~\eqref{eq:S2_even} provide rapidly convergent one–dimensional integrals that can be evaluated numerically with high precision for any $(x,z)$ in the physically relevant domain $0<x<1$, $0<z\le1$.

\section{Asymptotics of Vector Resonant Relaxation}
\label{app:asymp}

In this appendix we use the Legendre sums in Appendix~\ref{App:Legendre} to obtain the asymptotics of the VRR Hamiltonian \eqref{eq:H_VRR}. This Hamiltonian is
\begin{equation}
    H =
    -\frac{1}{2} \sum_{i,j=1}^{N}\sum_{\ell=0,2}^{\infty} \J_{ij\ell} P_{\ell}(\hat{\bm{L}}_i\cdot \hat{\bm{L}}_j)
    =-\frac{1}{2} \sum_{i,j=1}^{N}
    \sum_{\ell,m}   \frac{\J_{ij\ell}}{2\ell+1} Y_{\ell}^m(\hat{\bm{L}}_i)Y_{\ell}^{*m}( \hat{\bm{L}}_j)\,.
\end{equation}
As remarked above, the evolution was shown to be dominated by $\ell\rightarrow \infty$ for radially overlapping orbits. Here we determine the  asymptotic limit for large $\ell$ for both radially overlapping and non-overlapping orbits. 

\citet{Kocsis2015} has shown that asymptotically for $\ell\rightarrow\infty$ (see Eq. B68 therein)
\begin{equation}
\label{eq:J_ijl_asymp}
    \J_{ij\ell} = \frac{\J_{ij}}{\ell^2} ~~~{\rm where}~~~ 
    \J_{ij}=
    \left\{
        \begin{array}{ll}
        \dfrac{G m_{i} m_{j}}{\pi^2} 
    \dfrac{[(1+e_{\rm in})(1-e_{\rm out})]^{3/2}}{(e_{i}e_{j})^{1/2}} \dfrac{r_{\rm a,in}^{\ell}}{r_{\rm p,out}^{\ell+1}}  & {\rm if~non-overlapping}\,,  \\[2ex]
        \dfrac{4}{\pi^3} \dfrac{G m_{i} m_{j}}{a_{i} a_{j}} I_2 & {\rm if~overlapping}\,,
    \end{array}
    \right.
\end{equation}
where $I_2\equiv I^{(2)}(r_{\rm p,in},r_{\rm p,out},r_{\rm a,in},r_{\rm a,out})$ is an analytic function expressed with elliptic integrals in a closed form. 

For overlapping orbits the pairwise Hamiltonian simplifies asymptotically using the Legendre sums listed above as
\begin{equation}\label{eq:Hij_overlap}
    H_{ij} = -\sum_{\ell=1}^{\infty} \frac{\J_{ij}}{\ell^2}P_{\ell}(x) = \J_{ij}\left[-\frac{\pi^2}{6}+\chi(s)+\chi(c)+2\ln s \ln c\right]
    \approx \J_{ij}\left( b_1 + b_2 |\sin \theta|\right)
\end{equation}
where $x = \cos \theta = \L_i\cdot \L_j$, $s=\sqrt{(1-x)/2}=\sin(\theta/2)$, $c=\sqrt{(1+x)/2}=\cos(\theta/2)$, $b_1=-\pi^2/24$, $b_2=2\chi(2^{-1/2}) + \frac12 (\ln 2)^2 - \frac18 \pi^2$, and $\chi(z)=\sum_{n=0}^{\infty} z^{2n+1}/(2n+1)^2$ is the Legendre-chi function which may be expressed with the dilogarithm $\mathrm{Li}_2(z)=\sum_{n=1}^{\infty}z^{n}/n^2$ as $\chi(z)=\frac12[\mathrm{Li}_2(z)-\mathrm{Li}_2(-z)]$. The last approximate equality is accurate to better than $0.5\%$ for all $\theta$. Note that $b_1$ can be set to 0 in practice, that term does not affect the evolution. Equation~\eqref{eq:Hij_overlap} shows that for overlapping orbits, the VRR interaction energy is minimised for parallel or anti-parallel orbits (corresponding to stable equilibria) and it is maximised for perpendicular orbits (unstable equilibria) \citep{Takacs2018}.

The equations of motion take the form
\begin{equation}
    \frac{\mathrm{d}\L_i}{\mathrm{d}t} = \mathbf{\Omega}_i \times \L_i = - \sum_{j\ell} \frac{\J_{ij\ell}}{L_i L_j} P'_{\ell}(\Ln_i\cdot \Ln_j) \L_j \times \L_i
\end{equation}
where $L_i=m_i\sqrt{G\MSMBH a_i(1-e_i^2)}= G \MSMBH m_i a_i^{-1}(1-e_i^2)^{1/2}\omega_{{\rm orb},i}^{-1}$ and $\omega_{{\rm orb},i}=(G\MSMBH /a_i^3)^{1/2}$. The angular frequency is
\begin{equation}
    \mathbf{\Omega}_{i,j}  = - \sum_{\ell} \frac{\J_{ij\ell}}{L_i} P'_{\ell}(\Ln_i\cdot \Ln_j) \Ln_j 
     = -  k_{ij} \omega_{\rm orb,i}  \Ln_j         
     \sum_{\ell=2,4}^{\infty} \dfrac{P'_{\ell}(\Ln_i\cdot \Ln_j)}{\ell^2}  z^{\ell}
\end{equation}  
where
\begin{equation}     
    k_{ij}= \frac{\J_{ij}}{L_i\omega_i} =   
    \left\{
        \begin{array}{ll}
        \dfrac{m_{j}a_i}{\pi^2 \MSMBH r_{\rm p,out}} 
    \dfrac{[(1+e_{\rm in})(1-e_{\rm out})]^{3/2}}{(e_{i}e_{j})^{1/2}(1-e_i^2)^{1/2}}  & {\rm if~non-overlapping}\,  \\[2ex]
        \dfrac{4}{\pi^3} \dfrac{m_{j}}{\MSMBH} \dfrac{I_2}{a_j(1-e_i^2)^{1/2}} & {\rm if~overlapping}\,.
    \end{array}
    \right.
\end{equation}
and
\begin{equation}
    z = 
        \left\{
        \begin{array}{ll}
        \dfrac{r_{\rm a,in}}{r_{\rm p,out}} & {\rm if~nonoverlapping}\,,  \\[2ex]
        1 & {\rm if~overlapping}\,.
    \end{array}
    \right.
\end{equation}
Note that $z\leq 1$.

Thus, the general asymptotic expression valid for both overlapping ($z=1$) and non-overlapping orbits ($z<1$) is 
\begin{equation}
\label{eq:Omega_asymp_non-over}
    \mathbf{\Omega}_{i,j}  = -  k_{ij} \omega_{\rm orb,i} 
     \left\{
     \frac{1/2}{1-x}\ln\left[\frac{(1+X+z)(1+Y-z)}{4}\right]
        -\frac{1/2}{1+x}\ln\left[\frac{(1+X-z)(1+Y+z)}{4}\right]
     \right\}\Ln_j\,,         
\end{equation}  
where $x = \cos \theta = \L_i\cdot \L_j$, $X=(1-2 xz + z^2)^{1/2}$, $Y=(1+2 xz + z^2)^{1/2}$,
and for overlapping orbits this simplifies as
\begin{equation}
\label{eq:Omega_asymp_over}
    \mathbf{\Omega}_{i,j}  = -  k_{ij} \omega_{\rm orb,i} 
     \left\{\frac{\ln[(1+s)c]}{4s^2}
        -\frac{\ln[(1+c)s]}{4c^2}
        \right\}\Ln_j \approx -  \frac{1}{2}k_{ij} \omega_{\rm orb,i}\cot \theta\,\Ln_j \,.
\end{equation}  
The asymptotic expansion is expected to reproduce the exact evolution for the overlapping case for which the large $\ell$ modes have a dominant contribution over the small ones. In reality, this expression is expected to produce accurate results if augmented by corrections corresponding to the exact contributions of low $\ell$ terms, which are not properly reproduced with the large $\ell$ asymptotics, which we will derive in Appendix~\ref{app:EOM-corr} below.

\section{VRR energy for circular orbits and equations of motion}
\label{app:energy}
Building on the results in Appendix~\ref{app:asymp}, let us derive a uniformly valid approximation for the multipole sum in the VRR Hamiltonian -- that is, an asymptotic re-summation of the Legendre-polynomial sum over $\ell$ in equation \eqref{eq:H_VRR}. 

\begin{figure*} 
    \centering    \includegraphics[width=0.99\linewidth]{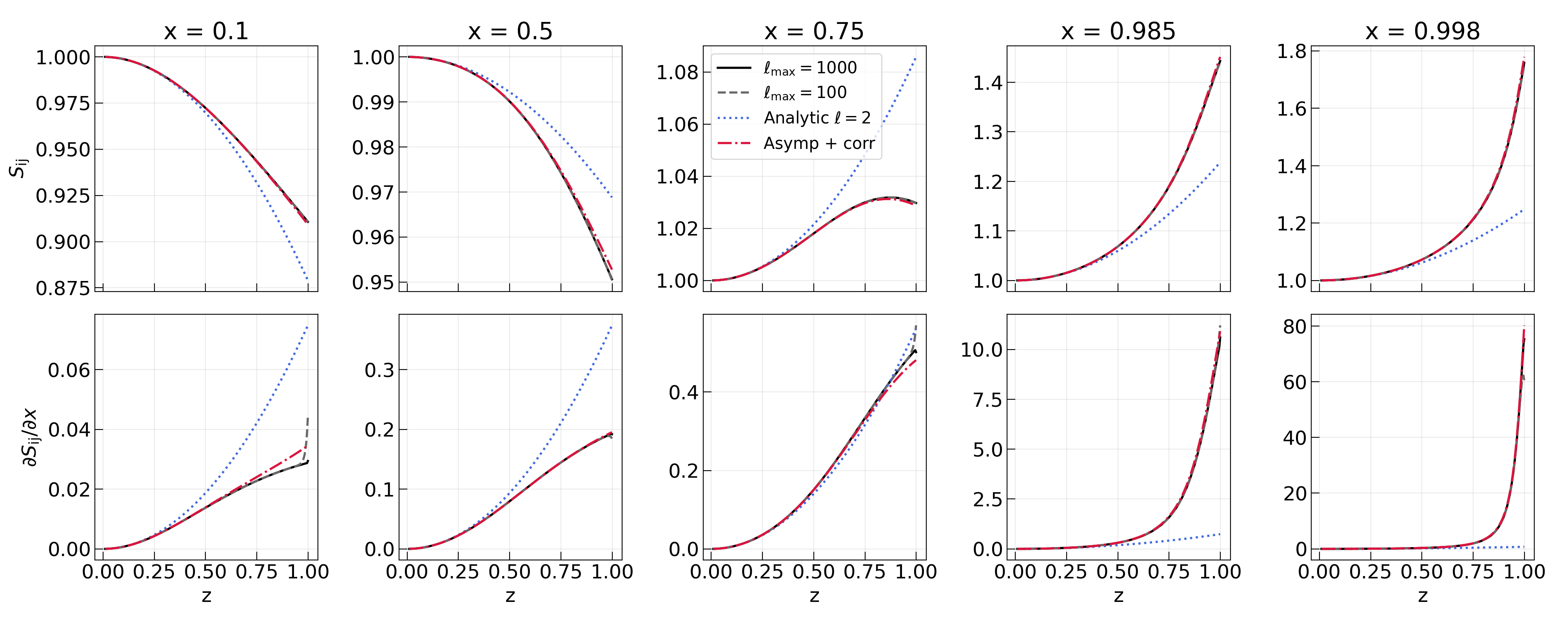}
    \caption{Infinite sum $\sum_{\ell=0}^{\infty}[P_{\ell}(0)]^2P_{\ell}(x)z^{\ell}$ (top row) and its derivative $\sum_{\ell=0}^{\infty}[P_{\ell}(0)]^2P'_{\ell}(x)z^{\ell}$ (bottom row) as functions of $z$ for given values of $x$. Solid and dashed black lines show numerical evaluation of the sums up to $\ell_\mathrm{max}=1000$ and 100, respectively. Blue dotted line corresponds to quadrupole approximation and dash-dotted red line shows asymptotic expansion with correction (Eq.~\ref{eq:S'_ij}). Clearly, the quadrupole approximation fails outside of $z\lapprox 0.3$, while the asymptotic expansion with correction is valid for arbitrary values of $z$.}
    \label{fig:S_ij}
\end{figure*}

To calculate the pairwise interaction energy for circular orbits we can write 
\begin{equation}
H_{ij} = \frac{ G m_i m_j}{r_{\Out}}S_{ij},
\end{equation}
where $S_{ij}$ is the infinite sum given by:
\begin{align}\label{eq:Scirc infinite sum}
    S_{ij}&=\sum_{\ell=0}^{\infty}[P_{\ell}(0)]^2P_{\ell}(x)z^{\ell},
\end{align}
and $z = r_{\rm in}/r_{\rm out}$. 
Below, we show how to find approximate closed-form representation of the sum above. Given the Legendre sums, listed in Appendix~\ref{App:Legendre}, we can write:
\begin{align}\label{eq:Scirc}
    S_{ij}&=\sum_{\ell=0}^{\infty}[P_{\ell}(0)]^2P_{\ell}(x)z^{\ell}
        =   
        \sum_{\ell=0}^{5}[P_{\ell}(0)]^2P_{\ell}(x)z^{\ell} +
    \frac{1}{\pi}\sum_{\ell=6}^{\infty}\frac{P_{2\ell}(x)}{\ell}z^{2\ell} 
    -    \frac{1}{4\pi}\sum_{\ell=6}^{\infty}\frac{P_{2\ell}(x)}{\ell^2} z^{2\ell}
    + \frac{1}{32\pi}\sum_{\ell=6}^{\infty}\frac{P_{2\ell}(x)}{\ell^3} z^{2\ell}
    +\cdots\,,
    \\
    &\approx 1 -
    \left(\frac{1}{\pi}-\frac{1}{4}\right) P_2(x) z^2 
    - \left(\frac1{2\pi}-\frac{9}{64}\right) P_4(x) z^4  
    +
    \frac{1}{\pi}\ln\left(\frac{2}{1-x z + X}\right)+\frac{1}{\pi}\ln\left(\frac{2}{1 + x z + Y}\right)
    \,,
\end{align}
where again, 
we have used
\begin{equation}
\sum_{\ell=1}^{\infty}\frac{P_{2\ell}(x)}{\ell} z^{2\ell} 
    = 
    \ln\left(\frac{2}{1-x z + X}\right)+\ln\left(\frac{2}{1 + x z + Y}\right)\,,
    \\
\end{equation}
and $x = \cos \theta = \L_i\cdot \L_j$, $X=(1-2 xz + z^2)^{1/2}$, and $Y=(1+2 xz + z^2)^{1/2}$.
This is accurate to better than $8\%$ for $z\leq 1-10^{-5}$ and almost all $x$ other than a narrow region (of width $\Delta x=0.03$) where $S_{ij}$ is close to zero, even if neglecting the unimportant constant term $1$. Note that we included only the first and second series on the right-hand side in Eq.~\eqref{eq:Scirc}; if we included the third one, the accuracy would improve to better than $1\%$ (except in a much narrower range where the result is close to zero). In the regions where the relative error is large -- where $S_{ij}$ is close to zero --  the absolute error is small of order $10^{-3}$, so that the approximation works well, there, too. 

For $x=1$ the sum can be computed for all $z<1$ as
\begin{align}
    S_{ij}&=\sum_{\ell=0}^{\infty}[P_{\ell}(0)]^2 z^{\ell}
        = \frac{2}{\pi}K(z)  
    \,,
\end{align}
where $K(k)$ is the complete elliptic integral of the first kind. This diverges near $x=z=1$ such that if $x=1$ and $z=1-\epsilon$ then
\begin{align}
    S_{ij}&=
        \frac{1}{\pi}\ln(\epsilon^{-1})+\ln 2 + \mathcal{O}(\epsilon)
    \,.
\end{align}

To get the precession rate, we take the derivative with respect to $x$, using $\partial X/\partial x = z/X$ and 
$\partial Y/\partial x = -z/Y$:
\begin{equation}
\begin{aligned}
S'_{ij} (x, z) = &\sum_{\ell=2}^{\infty}[P_{\ell}(0)]^2P'_{\ell}(x)z^{\ell} \approx
 -\left(\frac{1}{\pi}-\frac{1}{4}\right) P'_2(x)z^2  -\left(\frac1{2\pi}-\frac{9}{64}\right) P'_4(x)z^4 + 
\frac{z(1+X)}{\pi(1-xz+X)X}- \frac{z(1+Y)}{\pi(1+xz+Y)Y}.
\label{eq:S'_ij}
\end{aligned}
\end{equation} 
Note that $P'_{0}(x) = 0$ and $P_{1}(0) = 0$ so we can start the summation from $\ell = 2$. In Figure~\ref{fig:S_ij} we compare the exact evaluation of the sum on the left up to $\ell=100$ and $\ell=1000$ with analytical expressions in the quadrupolar approximation ($\ell=2$) and analytical expression given by Eq.~\eqref{eq:S'_ij} which is denoted in the figure as `Asymp + corr'. 

For $z$ close to unity, the asymptotic terms dominate the sum (last two terms on the right hand side of equation \eqref{eq:S'_ij}). Keeping only them and simplifying, gives an approximation for $S'_{ij}(\cos \theta,1)$, \emph{viz}.
\begin{equation}
S'_{ij} (\cos\theta, 1) \approx  \frac{2\cos\theta +\sqrt{2}\sin(\frac{\theta}{2} - \frac{\pi}{4})}{\pi \sin^2\theta},
\label{eq:S'_ij_asymp}
\end{equation}
which for nearly coplanar orbits reduces to
\begin{equation}
S'^{(\theta\ll1)}_{ij} (\cos\theta, 1) \approx \frac{1}{\pi \theta^2}.
\end{equation}
Thus, the precession rate for the nearly coplanar circular orbits with the same radii is:
\begin{equation}
\Omega_{i,j} = 2\pi\omega_i \frac{m_{j}}{\MSMBH} \frac{r_{i}}{r_{\Out}}S'_{ij} = 2\pi\omega_i \frac{m_{j}}{\MSMBH} \frac{1}{\pi \theta^2}.
\label{eq:omega_circ_asymp_app}
\end{equation}
This is equation \eqref{eq:omega_circ_asymp}, which we utilise throughout the paper to estimate the torques between the initially coupled angular momentum binaries (two nearly coplanar circular rings).

\subsection{Asymptotic equations of motion with corrections}
\label{app:EOM-corr}

Given this asymptotic expressions above, we can generalise the procedure described above and construct the exact equation of motion as
\begin{align}
\label{eq:EOM-exact}
\dot{\bm{L}}_i^{\mathrm{exact}}
&= - \sum_j \left[
      \Omega_{\mathrm{asymp}}
      + \sum_{\ell=2,4}^{\infty}
        \frac{\J_{ij\ell}^{\mathrm{exact}} - \J_{ij\ell}^{\mathrm{asymp}}}{L_i}\,
        P'_{\ell}\!\left(\cos\theta_{ij}\right)
    \right] \Ln_j \times \bm{L}_i
\\[4pt]
&\approx - \sum_j \left[
      \Omega_{\mathrm{asymp}}
      + \left(
            3\,\frac{J^\mathrm{exact}_{ij2} - \J_{ij2}^{\mathrm{asymp}}}{L_i}
            - \frac{15}{2}\,\frac{J^\mathrm{exact}_{ij4} - \J_{ij4}^{\mathrm{asymp}}}{L_i}
        \right)\!
        \cos\theta_{ij}
      + \left(
            \frac{35}{2}\,\frac{J^\mathrm{exact}_{ij4} - \J_{ij4}^{\mathrm{asymp}}}{L_i}
        \right)\!
        \cos^{3}\theta_{ij}
    \right] \Ln_j \times \bm{L}_i,
\end{align}
where $\Omega_{\mathrm{asymp}}$ is the asymptotic limit of the angular frequency, $J^\mathrm{exact}_{ij\ell}$ is the coupling coefficient given by Eq.~\eqref{eq:Jijl} and $\J_{ij\ell}^{\mathrm{asymp}}$ is its asymptotic limit. 
Applying this to circular orbits, using the first 2 terms (up to $\ell=4$), we recover the equation of motion:
\begin{align}
    \dot{\bm{L}}_i =  2\pi\omega_i \frac{m_{j}}{\MSMBH} \frac{r_{i}}{r_{\Out}}S'_{ij}(\cos{\theta}, r_{\In}/r_{\Out})\Ln_j\times \bm{L}_i,    
\end{align}
where $S'_{ij}$ is given by Eq.~\eqref{eq:S'_ij}.

For eccentric orbits, $\J_{ij\ell}^{\mathrm{asymp}} = {\ell^{-2}}{\J_{ij}}$ is given by Eq.~\eqref{eq:J_ijl_asymp}. $\Omega_\mathrm{asymp}$ is given by Eq.~\eqref{eq:Omega_asymp_non-over}, $J^\mathrm{exact}_{ij\ell}$ is given by Eq.~\eqref{eq:Jijl}, where $s_{ij\ell}$ is an analytic function. So it is straightforward to write the equation for this case. For overlapping orbits, $\Omega_\mathrm{asymp}$ is a simple expression given by Eq.~\eqref{eq:Omega_asymp_over}. So, we can write:
\begin{align}
\dot{\bm{L}}_i^{\mathrm{exact}}
\approx - \sum_j \left[k_{ij} \omega_{\rm orb,i}\cot \theta_{ij} 
    + \left[\frac{3\J_{ij(\ell=2)}^{\rm exact} + \frac{15}{2}\J_{ij(\ell=4)}^{\rm exact}}{L_i}-\left(\frac34 - \frac{15}{32} \right)k_{ij} \omega_{\rm orb,i}\right]  \cos\theta_{ij}
    + \left[\frac{\frac{35}{2}\J_{ij(\ell=4)}^{\rm exact}}{L_i}-\frac{35}{32} k_{ij} \omega_{\rm orb,i}\right]  \cos^{3}\theta_{ij} \right] \Ln_j\times \bm{L}_i     
\end{align}
Here $\J_{ij\ell}^{\rm exact}-\J_{ij\ell}^{\rm asymp}\propto \ell^{-3}$ for large $\ell$ which converges quickly, implying that this sum may be truncated at low orders. Taking the leading order term only, i.e. the asymptotic sum ($\propto\cot\theta$), is generally correct to within a $40\%$ error, including the first correction ($\ell=2$, $\propto\cos\theta$) is correct to $10\%$, and with the second term ($\ell=4$, $\propto\cos^3\theta$) it is correct to $2\%$. However for small angles the accuracy scales with $\sin \theta$. The expression $k_{ij}$ depends on the function $I^{(2)}$ (given below in Appendix \ref{app:I2}) and, in general case, can be evaluated numerically, but for the case of a circular ring embedded within an elliptical one, it simplifies analytically, as we show below. Additionally, $\J_{ij\ell}^{\rm exact}$ depends on the coefficient $s_{ij\ell}$ given by Eq.~\eqref{eq:s_ijl} which can be evaluated numerically, but simplifies for some particular cases (see Appendix B of \citealt{Kocsis2015}). And again, for the case of a circular ring within an elliptical, one can get rid of the outer integral.

\subsection{Endpoint expansion of \(I^{(2)}\) for \(c\to b^{+}\)}
\label{app:I2}
For a circular orbit embedded within an elliptical one, we can simplify $I^{(2)}(a,b,c,d)$, since in this case $b=c$.
Embedded orbits (rings) are a subclass of overlapping orbits in which one ring is radially embedded within the other, such that
$r_{p,1} < r_{p,2} < r_{a,2} < r_{a,1}$. If the embedded ring is circular, then $r_{p,2} = r_{a,2}$, corresponding to the limit
$c \to b$. In this case, the other ring must be elliptical.

We have
\begin{equation}
I^{(2)}(a,b,c,d)=
\int_{b}^{c}
\frac{r^{2}\,\mathrm{d}r}{
\sqrt{(r-a)(r-b)(c-r)(d-r)}}\,,
\qquad a<b<c<d .
\label{eq:I2_def}
\end{equation}
To evaluate the limit \(c\to b^{+}\) we introduce a small parameter
\(\epsilon>0\) such that \(c=b+\epsilon\) and set \(u=r-b\;(0\le u\le\epsilon)\):
\begin{equation}
I^{(2)}(a,b,b+\epsilon,d)=
\int_{0}^{\epsilon}
\frac{(b+u)^{2}\,\mathrm{d}u}{
\sqrt{(b+u-a)\,u\,(\epsilon-u)\,(d-b-u)}} .
\label{eq:I2_eps}
\end{equation}

For \(\epsilon\to0\) only the factor \(u(\epsilon-u)\) vanishes; expand the
others at \(u=0\):
\[
(b+u)^{2}=b^{2}+O(u),\qquad
b+u-a=b-a+O(u),\qquad
d-b-u=d-b+O(u).
\]
Keeping the leading terms gives
\begin{equation}
I^{(2)}(a,b,b+\epsilon,d)=
\frac{b^{2}}{\sqrt{(b-a)(d-b)}}
\int_{0}^{\epsilon}\frac{\mathrm{d}u}{\sqrt{u(\epsilon-u)}}
+O(\epsilon).
\label{eq:I2_leading}
\end{equation}

The remaining integral is elementary — substituting \(u=\epsilon\sin^{2}\phi\)
or using the Euler beta function:
\begin{equation}
\int_{0}^{\epsilon}\frac{\mathrm{d}u}{\sqrt{u(\epsilon-u)}}=\pi ,
\label{eq:beta_pi}
\end{equation}
independent of \(\epsilon\).

Thus, we arrive at the finite limit for an embedded circular orbit
\begin{equation}
\lim_{\epsilon\to0^{+}} I^{(2)}(a,b,b+\epsilon,d)=
\frac{\pi\,b^{2}}{\sqrt{(b-a)(d-b)}} .
\label{eq:I2_limit}
\end{equation}

\section{Numerical simulations}
\label{App:simulations}

We utilise 2 different codes: \nring (which solves equation of motion \ref{eq:EOM}, \citealt{Kocsis2015}) to check the analytical calculations, and \phicpu to run 4-body simulations which solve Newtonian equation of motion, which we describe briefly below.

\subsection{N-Ring}
\label{app:n-ring}

The \textsc{N{-}Ring} integrator advances the unit-vector angular momenta \(\bm{L}_i=L_i\hat{\bm{L}}_i\) of \(N\) stellar orbits under the VRR Hamiltonian, which is a sum of pairwise couplings. For each pair \((i,j)\) the double orbit-averaged (over the orbital period and apsidal precession period) equations of motion
\[
\frac{\mathrm d\bm{L}_i}{\mathrm dt}= -\sum_{\ell=2}^{\infty}
\frac{\mathcal{J}_{ij\ell}}{L_i L_j}\,P'_\ell\!\left(\hat{\bm{L}}_i\!\cdot\!\hat{\bm{L}}_j\right)\,
\bm{L}_j\times\bm{L}_i,\qquad
\frac{\mathrm d\bm{L}_j}{\mathrm dt}= -\,\frac{\mathrm d\bm{L}_i}{\mathrm dt}
\]
reduce, after the change of variables \(\bm{J}_{ij}=(\bm{L}_i+\bm{L}_j)/2\) and \(\bm{K}_{ij}=(\bm{L}_i-\bm{L}_j)/2\), to uniform precession of \(\bm{K}_{ij}\) about the fixed axis \(\bm{J}_{ij}\) with angular velocity
\[
\bm{\Omega}_{ij}= -2\sum_{\ell=2}^{\infty}\frac{{J}_{ij\ell}}{L_i L_j}
P'_\ell\left(\frac{\J_{ij}^2-K_{ij}^2}{L_i L_j}\right)\,\hat{\bm{J}}_{ij}.
\]
Because this pairwise evolution is analytic -- a rigid rotation by the angle \(\Omega_{ij}\,\Delta t\) -- each pair update exactly preserves phase–space volume and integrals of motion.

{\nring} constructs a global symplectic map by composing all \(N(N-1)/2\) pair operators \(\mathcal{O}_{ij}(\Delta t)=\exp(\Delta t\,\mathcal{G}_{ij})\) in a time-reversible order, yielding a second-order scheme that can be promoted to eighth order through triple-jump factorisation. A multilevel block-time-step refinement decreases the step size only for the innermost or most strongly-coupled stars, while maintaining symplecticity, and a binary-tree tiling lets non-overlapping pairs update simultaneously, giving \(\mathcal{O}(N^2/P)+2P\) scaling on \(P\) processors.  The result is an exactly orbital-energy-conserving and angular-momentum-vector conserving integrator whose dominant cost is \(N^2\) pair rotations, but whose accuracy can be tuned by the time-step hierarchy and the chosen operator-splitting order. Details on {\nring} may be found in \cite{Kocsis2015}.

\subsection{$\phi$-CPU}
\label{app:phi-cpu}

We use a modified version of the $\phi$-GPU code\footnote{$N$-body code $\phi$-GPU: \\~\url{https://github.com/berczik/phi-GPU-mole}} \citep{Berczik2011,BSW2013}, optimised to CPU-bound tasks, which we therefore refer to as \phicpu. The code solves the 4-body problem: the central SMBH, an IMBH and 2 stars. The coordinate system is chosen such that the SMBH is at the origin, and the IMBH is initially in the XY-plane. Both SMBH and IMBH have no softening ($\eps_\mathrm{bh} =0$) and reduced time-steps, while stars have a softening length of $\eps_\mathrm{ss} = 10^{-4}$ (in code units). The equation of motion is:
\begin{equation}\label{eq:eom_plummer}
\ddot{\mathbf{r}}_i =
-\,G\sum_{j\neq i} m_j \frac{\mathbf{r}_{ij}}{(r_{ij}^{2}+\epsilon_{ij}^{2})^{3/2}}
-\,G\,M_{\mathrm{pl}} \frac{\mathbf{r}_i}{(r_i^{2}+a_{\mathrm{pl}}^{2})^{3/2}},
\end{equation}
where $\mathbf{r}_{ij}=\mathbf{r}_i-\mathbf{r}_j$ and $\epsilon_{ij}$ is the softening length for the pair $(i,j)$. 
Internal units are set to $G=\MSMBH=R_{\mathrm{out}}=1$, with $R_{\mathrm{out}}=0.5\,\mathrm{pc}$ and $\MSMBH=4\times10^{6}\,M_\odot$ for conversion to physical scales.

\bsp	
\label{lastpage}
\end{document}